\documentclass[preprint,5p,times,twoecolumn]{elsarticle}
\usepackage{graphicx,amssymb}
\usepackage{ifpdf}
\usepackage{xspace}
\usepackage{multirow}
\usepackage{setspace}
\usepackage{pbox}
\usepackage{wrapfig}
\usepackage{rotating}
\usepackage{xcolor}
\usepackage{setspace}

\usepackage{etoolbox}

\definecolor{darkgreen}{HTML}{4CBB17}
\definecolor{blue}{HTML}{2D8BAE}
\newcommand{\green}[1]{#1}
\newcommand{\blue}[1]{#1}

\makeatletter

\patchcmd{\ps@pprintTitle}{\footnotesize\itshape
       Preprint submitted to \ifx\@journal\@empty Elsevier
       \else\@journal\fi\hfill\today}{This article is accepted for publication and will appear in the Journal of Biomedical Informatics.\hfill}{}{}

\def\elsartstyle{
    \def\normalsize{\@setfontsize\normalsize\@xiipt{14.5}}
    \def\small{\@setfontsize\small\@xipt{13.6}}
    \let\footnotesize=\small
    \def\large{\@setfontsize\large\@xivpt{18}}
    \def\Large{\@setfontsize\Large\@xviipt{22}}
    \skip\@mpfootins = 18\p@ \@plus 2\p@
    \normalsize
}
\@ifundefined{square}{}{}
\makeatother

\graphicspath{{jwsGraphics/}}
\usepackage[center]{subfigure}

\newcommand{\subfiguresizehoz}{0.32}

\newcommand{\subfiguresizecontent}{0.38}
\newcommand{\chao}{ChAO\xspace}
\newcommand{\pro}{Prot\'eg\'e\xspace}
\newcommand{\cprot}{Collaborative Prot\'eg\'e\xspace}
\newcommand{\wpro}{Web\-Prot\'eg\'e\xspace}
\newcommand{\icat}{iCAT\xspace}
\newcommand{\numdatasets}{five\xspace}

\newcommand{\captionncit}{National Cancer Institute Thesaurus (NCIt)\xspace}
\newcommand{\captionicd}{International Classification of Diseases (\mbox{ICD-11})\xspace}
\newcommand{\captionictm}{International Classification of Traditional Medicine (ICTM)\xspace}

\newcommand{\captionopl}{Ontology for Parasite Lifecycle (OPL)\xspace}
\newcommand{\captionbro}{Biomedical Resource Ontology (BRO)\xspace}

\usepackage{enumitem}
\usepackage{hyperref}

\begin{document}

\begin{frontmatter}

\title{Discovering Beaten Paths in\\Collaborative Ontology-Engineering Projects\\\blue{using Markov Chains}}

\author[iicm]{Simon Walk\corref{cor1}}
\author[gesis]{Philipp Singer}
\author[gesis,kob]{Markus Strohmaier}
\author[su]{Tania Tudorache}
\author[su]{Mark A. Musen}
\author[su]{Natalya F. Noy}

\medskip

\cortext[cor1]{Corresponding author (\url{simon.walk@tugraz.at})}
\address[iicm]{Institute for Information Systems and Computer Media, Graz University of Technology, Austria}
\address[gesis]{GESIS - Leibniz Institute for the Social Sciences, Cologne, Germany}
\address[kob]{Dept. of Computer Science, University of Koblenz-Landau, Germany}
\address[su]{Stanford Center for Biomedical Informatics Research, Stanford University, USA}

\begin{abstract}
Biomedical taxonomies, thesauri and ontologies in the form of the International Classification of Diseases as a taxonomy or the National Cancer Institute Thesaurus as \blue{an OWL-based ontology}, play a critical role in acquiring, representing and processing information about human health. With increasing adoption and relevance, biomedical ontologies have also significantly increased in size. For example,  the 11$^{\textnormal{th}}$ revision of the International Classification of Diseases, which is currently under active development by the World Health Organization
contains nearly $50,000$ classes representing a vast variety of different diseases and causes of death. This evolution in terms of size was accompanied by an evolution in the way ontologies are engineered. Because no single individual has the expertise to develop such large-scale ontologies, ontology-engineering projects have evolved from small-scale efforts involving just a few domain experts to large-scale projects that require effective collaboration between dozens or even hundreds of experts, practitioners and other stakeholders. Understanding the way these different stakeholders collaborate will enable us to improve editing environments that support such collaborations. In this paper, we uncover how large ontology-engineering projects, such as the International Classification of Diseases in its 11$^{\textnormal{th}}$ revision, unfold by analyzing usage logs of five different biomedical ontology-engineering projects of varying sizes and  scopes \blue{using Markov chains}. We discover intriguing interaction patterns (e.g., which properties users frequently change after specific given ones) that suggest that large collaborative ontology-engineering projects are governed by a few general principles that determine and drive development. From our analysis, we identify commonalities and differences between \blue{different} projects that have implications for project managers, ontology editors, developers and contributors working on collaborative ontology-engineering projects and tools in the biomedical domain.

\end{abstract}


\end{frontmatter}

\section{Introduction}
\label{intro}

Today, biomedical ontologies play a critical role in acquiring, representing and processing information about human health. For example, the International Classification of Diseases (ICD) is \green{a taxonomy that is} used in more than 100 countries to encode patient diseases, to compile health-related statistics and to collect health-related spending statistics. Similarly, the National Cancer Institute's Thesaurus (NCIt) represents an important \blue{OWL-based} vocabulary for classifying cancer and cancer-related terms.

With their increase in relevance, biomedical taxonomies, thesauri and ontologies have also significantly increased in size to cover new findings and to extend and complement their original areas of application. For example, the 11$^{\textnormal{th}}$ revision of the International Classification of Diseases (ICD-11), currently under active development by the World Health Organization (WHO),  consists of nearly $50,000$ classes representing a vast variety of different diseases and causes of death. \green{In contrast to previous revisions, the foundation component of ICD-11 is implemented as an OWL ontology with a broader scope than previous ICD revisions.}

This  growth was accompanied by a need to adapt the way these ontologies are engineered as no single individual or small group of domain experts have the expertise to develop such large-scale ontologies. New tools and processes have to be developed in order to coordinate, augment and manage collaboration between the dozens or hundreds of experts, practitioners and stakeholders when engineering an ontology.

Understanding the ways in which such a large number of participants -- e.g., more than $100$ experts contribute to ICD-11 -- collaborate with one another when creating a structured knowledge representation is a prerequisite for quality control and effective tool support.

\textbf{Objectives:} Consequently, we  aim at understanding how large collaborative ontology-engineering projects such as ICD-11 unfold. 
\green{In particular, we want to investigate if we can identify usage patterns in the change-logs of collaborative ontology-engineering projects?} We approach this problem by analyzing patterns in usage logs of five biomedical ontology-engineering projects of varying sizes and scopes. \green{For this analysis we employ Markov chain models for investigating and modeling sequential interaction paths (c.f. Section~\ref{sub:sequential paths}). Such paths are represented by chronologically ordered lists of interactions within the underlying ontology for (a) a single user or (b) a single class (see Figure~\ref{fig:seq_pat_sample}). For example, we study sequences of properties that were either changed by (a) \emph{a single user} on any class or (b) \emph{a single class} by any user in an ontology over time. \blue{For example, as depicted in Figure~\ref{fig:seq_pat_sample}, a sequential property path for a single user (user-based) consists of a chronologically ordered list of all properties (e.g., \emph{title}, \emph{definition} etc.), which have been changed by that user on any class, while a sequential property path for a single class (class-based) consists of a chronologically ordered list of properties that were changed on that class by any user.} Instead of only modeling sequences for single users or classes, our data contains a set of paths; e.g., each path in the dataset consists of sequences of properties whose value has been changed by a single user over time. This allows us to tap into accumulated patterns. Concretely, we are interested in studying emerging patterns of subsequent steps in such sequential paths -- e.g., which properties do users frequently change after a specific given property.}

The analyzed datasets range from large-scale datasets such as ICD-11 to smaller ones such as the Ontology for Parasite Lifecycle (OPL). Given the differences of our datasets in a number of salient characteristics, we investigate if specific patterns can be found across all or only in certain biomedical ontology-engineering projects. \green{Furthermore, we investigate and discuss features of these projects that potentially affect observed patterns, which can only be found in specific datasets.} This analysis can be seen as a stepping stone for collaborative ontology-engineering project managers to devise infrastructures and tool support to augment collaborative ontology engineering.

\begin{figure*}[!th]
\centering
{\includegraphics[width=\linewidth]{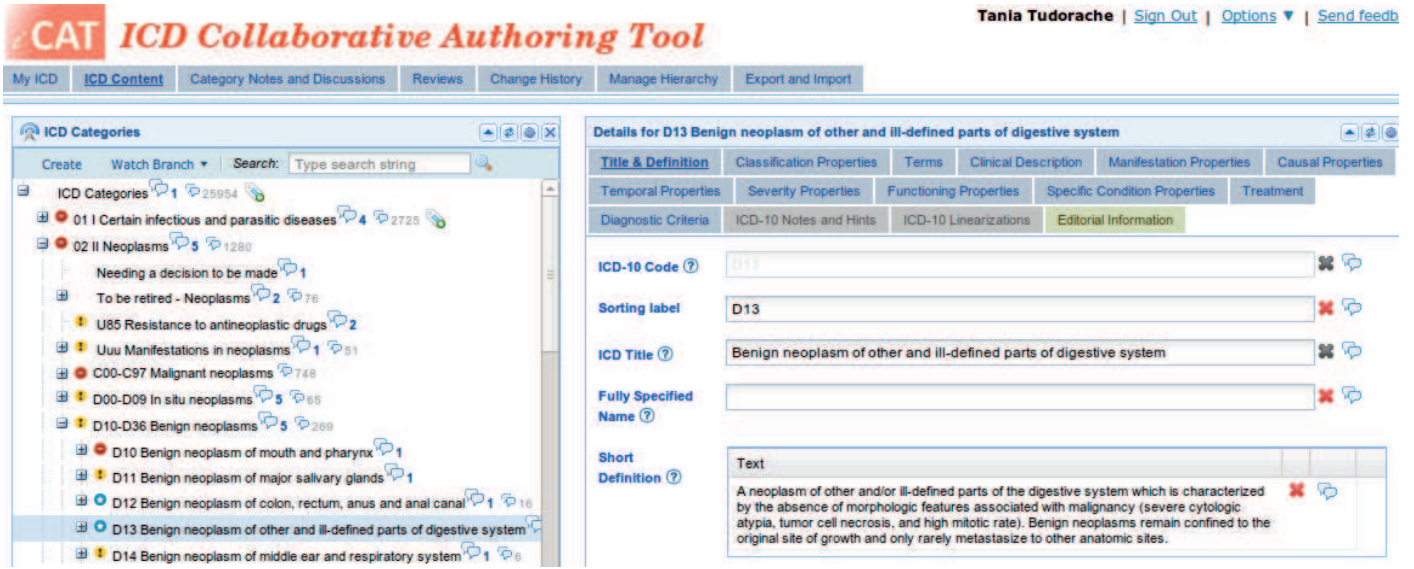}}
\caption{A screenshot of iCAT, a custom tailored, web-based version of \wpro, developed for the collaborative engineering of ICD-11. The left part of the interface visualizes the ICD-11 class hierarchy, the class titles, the number of annotations each class has received (speech bubbles) and its overall progress (color and symbol before the class title). The right part of the interface shows the different user-interface sections (e.g, \emph{Title \& Definition} or \emph{Classification Properties}), listing all properties and property values for each class. 
}
\label{fig:icat}
\end{figure*}

\textbf{Contributions:} We present new insights on social interactions and editing patterns that suggest that large collaborative ontology-engineering projects are governed by a few general principles that determine and drive development. 
Specifically, our results indicate that general edit patterns can be found in all investigated datasets, even though they (i) represent different projects with different goals, (ii) use variations of the same ontology-editors and tools for the engineering process and (iii) differ in the way the projects are coordinated. 

To the best of our knowledge, the work presented in this paper represents the most fine-grained and comprehensive study of patterns in large-scale collaborative ontology-engineering projects in the domain of biomedicine. In addition, our analysis is conducted across \numdatasets datasets of different sizes, which have been developed using different versions of \cprot (Table~\ref{tab:dataset details}).

\section{Collaborative ontology engineering}
\label{sub:collaborative ontology-engineering}

According to \citet{Gruber-A-1993,borst1997ceo,studer1998} an ontology is an explicit specification of a shared conceptualization. In particular, this definition refers to a machine-readable construct (the formalization) that represents an abstraction of the real world (the shared conceptualization), which is especially important in the field of computer science as it allows a computer (among other things) to ``understand'' relationships between entities and objects that are modeled in an ontology.

Collaborative ontology engineering is a new field of research with many new problems, risks and challenges that we must first identify and then address.
In general, contributors of collaborative ontology-engineering projects, similar to traditional collaborative online production systems\footnote{\green{Note that the term traditional online production systems refers to online platforms that have users collaborate in engineering digital goods, opposed to a structured knowledge base that is the result of collaborative ontology-engineering.}} (e.g., Wikipedia), engage remotely (e.g., via the internet or a client--server architecture) in the development process to create and maintain an ontology.  
As an ontology represents a formalized and abstract representation of a specific domain, disagreements between authors on certain subjects can occur. Similar to face-to-face meetings, these collaborative ontology-engineering projects need tools that augment collaboration and help contributors in reaching consensus when modeling topics of the real world.

Indeed, the majority of the literature about collaborative ontology engineering sets its focus on surveying, finding and defining requirements for the tools used in these projects \cite{conf/aaaiss/NoyT08, Groza:2013:CSA:2435464.2435766}.  

The Semantic Web community has developed a number of tools aimed at supporting the collaborative development of ontologies. For example, Semantic MediaWikis \cite{krotzsch2006sm} and its derivatives \cite{Ontowiki.ISWC,ghidini2009moki,schandl2010poolparty} add semantic, ontology modeling and collaborative features to traditional MediaWiki systems.

\pro, and its extensions for collaborative development, such as \wpro and iCAT~\cite{WebProtege.SWJ} (see Figure~\ref{fig:icat} for a screenshot of the iCAT ontology-editor interface) are  prominent stand-alone tools that are used by a large community worldwide to develop ontologies in a variety of different projects. Both \wpro and \cprot provide a robust and scalable environment for collaboration and are used in several large-scale projects, including the development of ICD-11~\cite{ICD.ISWC}.

\citet{poeschko-aaai12} and \citet{walk-ijswis} have created \emph{PragmatiX}, a tool to visualize and analyze a collaboratively engineered ontology and aspects of its history and the engineering process, providing quantitative insights into the ongoing collaborative development processes. 

\citet{conf/kcap/FalconerTN11} investigated the change-logs of collaborative ontology-engineering projects, showing that users exhibit specific roles, which can be used to group and classify users, when contributing to the ontology.
\citet{10.1371/journal.pcbi.1002630} investigated whether the location and specific structural features can be used to determine if and where the next change is going to occur in the Gene Ontology\footnote{\url{http://www.geneontology.org}}.

\green{Goncalves et. al \cite{Goncalves:2011:AEN:2190647.2190839, gonccalves2011facilitating, Goncalves:2011:CLD:2063576.2063797} performed an analysis of different versions of ontologies by applying and categorizing \emph{Diff} algorithms, with the goal of categorizing the differences between consecutive and chronologically ordered versions of the ontologies. Furthermore, they conducted reasoner performance tests and identified factors that potentially increase reasoner performance. For the analysis presented in this paper we were able to rely on \chao \cite{noy2006framework}, which is a change-log provided by \pro and its derivatives that already provides us with detailed and unambiguous logs of changes for the investigated ontologies.

In a similar context \citet{grau2007just, Grau:2007:LFM:1625275.1625322} proposed a logical framework for modularity of ontologies and a definition of what is to be considered as an ontology module. In general, an ontology module can be used to extract the meaning of a specified set of terms from an ontology. Extracting the right amount of information is especially important for the topic of ontology reuse. According to Grau et al. modularity also represents a crucial factor in collaborative ontology-engineering environments as modular representations of ontologies are easier to understand, to extend and to reuse, similar to modularity in software engineering projects.

\citet{Mikroyannidi:2011:IRO:2063016.2063045} investigated the detection and use of (design) patterns in the content of an ontology, using a clustering approach. In contrast to Mikroyannidi et al., our analysis focuses on the detection of sequential patterns in interaction data rather than content.}

\citet{j.websem333} investigated the hidden social dynamics that take place in collaborative ontology-engineering projects from the biomedical domain and provides new metrics to quantify various aspects of the collaborative engineering processes. \citet{wang2013analysis} have used association-rule mining  to analyze user editing patterns in collaborative ontology-engineering projects. The approach presented in this paper uses Markov chains to extract much more fine grained user-interaction patterns incorporating a variable number of historic editing information.

The only requirement to perform the  pattern analysis that we present in this paper is the availability of a structured log of changes that can be mapped to the underlying ontology. The majority of the discussed collaborative ontology-engineering environments provide such a log, allowing for a similar analysis. For example, the Semantic MediaWikis store all the changes to the articles, and thus the ontology, allowing to expand the application of Markov chains to analyze sequential patterns as shown in this paper.

\section{Materials \& methods}
\label{materials and methods}

For the analysis conducted in this paper we concentrated our efforts on five  ontology-engineering projects in the biomedical domain. Each of the projects (i) has at least two users who contributed to the project, (ii) provides a structured log of changes and (iii) represents knowledge  from the biomedical domain.
In Section~\ref{sub:datasets} we provide a brief history for each dataset and in Section~\ref{sub:sequential paths} we describe the sequential path analysis. To aid readers in understanding the analyses conducted in this paper and its implications we provide a very brief overview of Markov chains and the involved model selection methodology in Section~\ref{sub:markov chain model selection}. 

\begin{table*}[t!]
\center
\footnotesize
\caption{Detailed information of the datasets used for the sequential pattern analysis to extract beaten paths in collaborative ontology-engineering projects.}
\begin{tabular}{| c | l | c | c | c | c | c |}
\cline{3-7}
\multicolumn{2}{c|}{} & ICD-11 & ICTM & NCIt & BRO & OPL \\\hline
\multirow{3}{*}{Ontology} & classes & 48,771 & 1,506 & 102,865 & 528 & 393 \\
 & changes & 439,229 & 67,522 & 294,471 & 2,507 & 1,993\\
& DL expressivity & $\mathcal{SHOIN}$(\textbf{D}) & $\mathcal{SHOIN}$(\textbf{D}) & $\mathcal{SH}$ & $\mathcal{SHIF}$(\textbf{D}) & $\mathcal{SHOIF}$ \\
\hline\hline
\multirow{1}{*}{Editor} & tool & iCAT & iCAT-TM & \cprot & \wpro & \cprot \\

\hline\hline
\multirow{2}{*}{Users} & users & 109 & 27 & 17 & 5 & 3 \\
& bots (changes) & 1 (935) & 1 (1) & 0 (0) & 0 (0) & 0 (0)\\
\hline\hline
\multirow{2}{*}{Duration} & first change & 18.11.2009 & 02.02.2011 & 01.06.2010 & 12.02.2010 & 09.06.2011 \\
& last change & 29.08.2013 & 17.7.2013 & 19.08.2013 & 06.03.2010 & 23.09.2011\\
& observation period (ca.) & 4 years & 2.5 years & 3 years & 1 month & 3 months\\
  \hline
\end{tabular}

\label{tab:dataset details}
\end{table*}

\subsection{Datasets}
\label{sub:datasets}

Table~\ref{tab:dataset details} lists the detailed features and observation periods for the following five datasets that we used in our analysis. 
\green{All datasets have been created either with \wpro or special versions of \wpro. To be able to conduct the pattern detection analysis for a different dataset, there is only one requirement that needs to be satisfied: The availability of a change-log that can be mapped onto the ontology so that changes can be associated with users and classes without ambiguity.}

\green{The DL expressivity \blue{\cite{Staab:2009:HO:1655829, Baader:2003:DLH:885746}} of the five datasets is added to Table~\ref{tab:dataset details} to highlight that the investigated ontologies exhibit different strategies regarding their OWL-DL expressivity. As all levels of expressivity shown in Table~\ref{tab:dataset details} allow for the definition and assignment of properties and classes, they do not influence the conducted pattern detection analyses. Also, in the case of \wpro and its derivatives, the data used for the pattern detection analysis can be extracted from the change-logs, allowing us to prevent parsing and extracting values from OWL directly.}

\textbf{The International Classification of Diseases (ICD)}\footnote{\url{http://www.who.int/classifications/icd/en/}} is the international standard for diagnostic classification used to encode information relevant to epidemiology, health management, and clinical use in over 100 United Nations countries. The World Health Organization (WHO) develops ICD, and publishes new revisions of the classification every decade or more. \green{The current revision in use is \mbox{ICD-10}, a taxonomy that contains over $15,000$ classes.} The 11th revision of ICD,\footnote{\url{http://www.who.int/classifications/icd/ICDRevision/}} \mbox{\textbf{ICD-11}}, is currently taking place and brings two major changes with respect to previous revisions. First, \mbox{ICD-11}\green{'s foundation component is developed as an OWL ontology} using a much richer representation formalism than previous revisions. ICD-11 contains very detailed descriptions of several aspects of diseases, mostly represented as properties in the ontology. Second, the development of ICD-11 takes place in a Web-based collaborative environment, called iCAT (see Figure~\ref{fig:icat}), which allows domain experts around the world to contribute and review the ontology online. \mbox{ICD-11} is planned to be finalized in May 2017.

\textbf{The International Classification of Traditional Medicine (ICTM)} is a WHO led project that aimed to produce an international
standard terminology and classification for diagnoses and interventions in Traditional Medicine.\footnote{\url{http://tinyurl.com/ictmbulletin}} ICTM, \green{similarly to ICD-11, is implements an OWL based ontology as foundation component,} which tries to unify the knowledge from the traditional medicine practices from China, Japan and Korea. Its content is authored in 4 languages: English, Chinese, Japanese and Korean. More than 20 domain experts from the three countries developed ICTM using a customized version of the iCAT system, called \mbox{iCAT-TM}. The development of ICTM was stopped in 2012, and a subset of ICTM is also included as a branch in the \mbox{ICD-11} ontology.\footnote{The ICD-11 dataset used in our analysis did not include the ICTM branch.}

\textbf{The National Cancer Institute's Thesaurus (NCIt)}~\cite{NCIT:07} has over $100,000$ classes and has been in development for more than a decade. It is a reference vocabulary covering areas for clinical care, translational, basic research, and cancer biology. A multidisciplinary team of editors works to edit and update the terminology based on their respective areas of expertise, following a well-defined workflow.
A lead editor reviews all changes made by the editors. The lead editor accepts or rejects the changes and publishes a new version of the NCI Thesaurus. The NCI Thesaurus is \blue{, at its core,} an OWL ontology, which uses many OWL primitives such as defined classes and restrictions. \blue{It was named thesaurus due to historical reasons, however fully conforms to OWL semantics, thus represents an actual ontology.}

\textbf{The Biomedical Resource Ontology (BRO)} originated in the Biositemaps project,\footnote{\url{http://biositemaps.ncbcs.org}} an initiative of the Biositemaps Working Group of the NIH National Centers for Biomedical Computing~\cite{broJBI}.
Biositemaps is a mechanism for researchers working in biomedicine to publish metadata about biomedical data, tools, and services. Applications can then aggregate this information for tasks such as semantic search. BRO is the enabling technology used in Biositemaps; a controlled terminology for describing the resource types, areas of research, and activity of a biomedical related resource. BRO was developed by a small group of editors, who use a Web-based interface (\wpro) to modify the ontology and to carry out discussions to reach consensus on their modeling choices.

\textbf{The Ontology for Parasite Lifecycle (OPL)} models the life cycle of the \textit{T.cruzi}, a protozoan parasite, which is responsible for a number of human diseases. OPL is an OWL ontology that extends several other OWL ontologies. It uses many OWL constructs such as restrictions and defined classes. Several users from different institutions collaborate on OPL development. This ontology is much smaller and has far fewer users than NCIt, \mbox{ICD-11}, or ICTM.

\subsection{Sequential interaction paths}
\label{sub:sequential paths}

\begin{figure}[!ht]
\centering
{\includegraphics[width=0.7\linewidth]{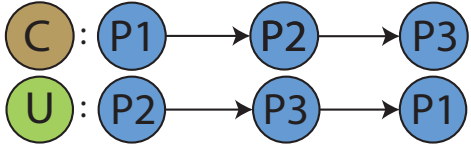}}
\caption{The top row of the figure depicts an exemplary \textbf{class-based} sequential property path ($P1$ to $P3$) for class $C$. This means that for class $C$ the property $P1$ was changed first, then property $P2$ and most recently changed was property $P3$. The bottom row of the figure depicts the sequential property path ($P1$ to $P3$), however this time for a user $U$ (\textbf{user-based}). Analogously, user $U$ has first changed $P2$, continued to change property $P3$ and most recently changed $P1$.}
\label{fig:seq_pat_sample}
\end{figure}

For our sequential pattern analysis we analyze three different kinds of paths, which all represent interactions with the underlying ontology. A sequential path is represented by the chronologically ordered list of extracted interactions for either a single user or a single class (see Figure~\ref{fig:seq_pat_sample}). For example, a sequential property path for a single user (user-based) consists of a chronologically ordered list of all properties (e.g., \emph{title}, \emph{definition} etc.), which have been changed by that user on any class, while a sequential property path for a single class (class-based) consists of a chronologically ordered list of properties that were changed on that class by any user.

\emph{User-sequence paths:}
First, we analyze activity patterns within the collaborative ontology-engineering project. This means that we analyze sequences of users who change a class. 
We want to detect and describe the different sequential patterns (the structure) that can be extracted from the change-logs of the investigated collaborative ontology-engineering projects.

\emph{Structural paths:} Analogously to the User-Sequence Paths, we investigate edit-strategies, such as \emph{bottom-up} or \emph{top-down} development, that users follow.
Is it possible to detect common patterns of which depth level a user frequently contributes to after a given current depth level?
In addition to development-strategies, we look at the relationships (e.g., parent, child, sibling, etc.) between the current and the next class a user is going to contribute to.

\emph{Property paths:} On a content-based level, we investigate the series of property-changes users perform on. 
In particular, we want to identify common successive property-changes -- i.e., which properties \emph{users} (user-based) regularly change consecutively and which properties are changed back-to-back for \emph{classes} (class-based).

\subsection{Markov chain models}
\label{sub:markov chain model selection}

For the analysis conducted in this paper we are adopting the methodology presented by \citet{singer_mc} and mapped to collaborative ontology-engineering change logs by \citet{walk_dev_mc} to detect sequential patterns identified in and extracted from change-logs of collaborative ontology-engineering projects.

For a better understanding of the collected results, we will provide a short description of Markov chains. For an in-depth description of our methodology  we point to \citet{singer_mc, walk_dev_mc}.

\green{
In general, Markov chain models are used for stochastically modeling transitions between states on a given state space.
In our case, a Markov chain consists of a finite \emph{state-space} (e.g., properties that a user edits over time; see Section~\ref{sub:sequential paths}) and the corresponding \emph{transition probabilities} (e.g., the probability of changing property j after property i) between these states.
Markov chain models are usually described as memoryless which means that the next state in a sequences only depends on the current one and not on a sequence of preceding ones (also known as Markovian property). Hence, this property defines serial dependence between adjacent nodes in trajectories -- this is where the term "chain" comes from. Such a model is usually called a \emph{first-order} or \emph{memoryless} model.

As we are interested in modeling sequential interaction paths of collaborative ontology-engineering projects (see Section~\ref{sub:sequential paths}), we fit a Markov chain model on such sequences $D=(x_1, x_2, ..., x_n)$ with states from a finite set $S$. Then, we can write the Markovian property as:

\begin{equation}
 P(x_{n+1}|x_1, x_2, ..., x_n)=P(x_{n+1}|x_n)
\end{equation}

After the model fitting on the data, a Markov chain model is usually represented via a stochastic transition matrix
$P$ with elements $p_{ij}=P(x_j|x_i)$ where it holds that for all $i$:

\begin{equation}
 \sum_j p_{ij} = 1
\end{equation}

For our analysis, we will make use of these transition probabilities to identify likely transitions for a variety of different states.\footnote{Note that throughout this article we usually refer to the entities modeled (i.e., interactions) instead of states. However, we speak about transition probabilities between these entities as we derive them directly from the resulting model transition matrix.} For example, if we fit the Markov chain model on sequential property paths for users (see Section~\ref{sub:sequential paths}), element $p_{ij}$ of the transition matrix would tell us the probability that users change property j right after i (e.g., in 60\% of all cases). By now, e.g., looking for the highest transition probabilities from state i to all other states of $S$, we can identify potential high-frequent patterns in our data.}

\section{Results}
\label{sequential pattern analysis}

\subsection{User-sequence paths}
\label{sub:activity paths}

In the \emph{User-Sequence Paths} analysis we investigate patterns emerging when looking at sequences of users who contribute to a class of an ontology. Hence, given a sequence of $n$ contributors for a class over time, we identify consecutive users who edit the class (e.g., user Y frequently contribute to a class after user X).

\begin{figure*}[!ht]
\centering
\subfigure[\captionicd]{\label{fig:ucsl:a}\includegraphics[width=\subfiguresizehoz\linewidth]{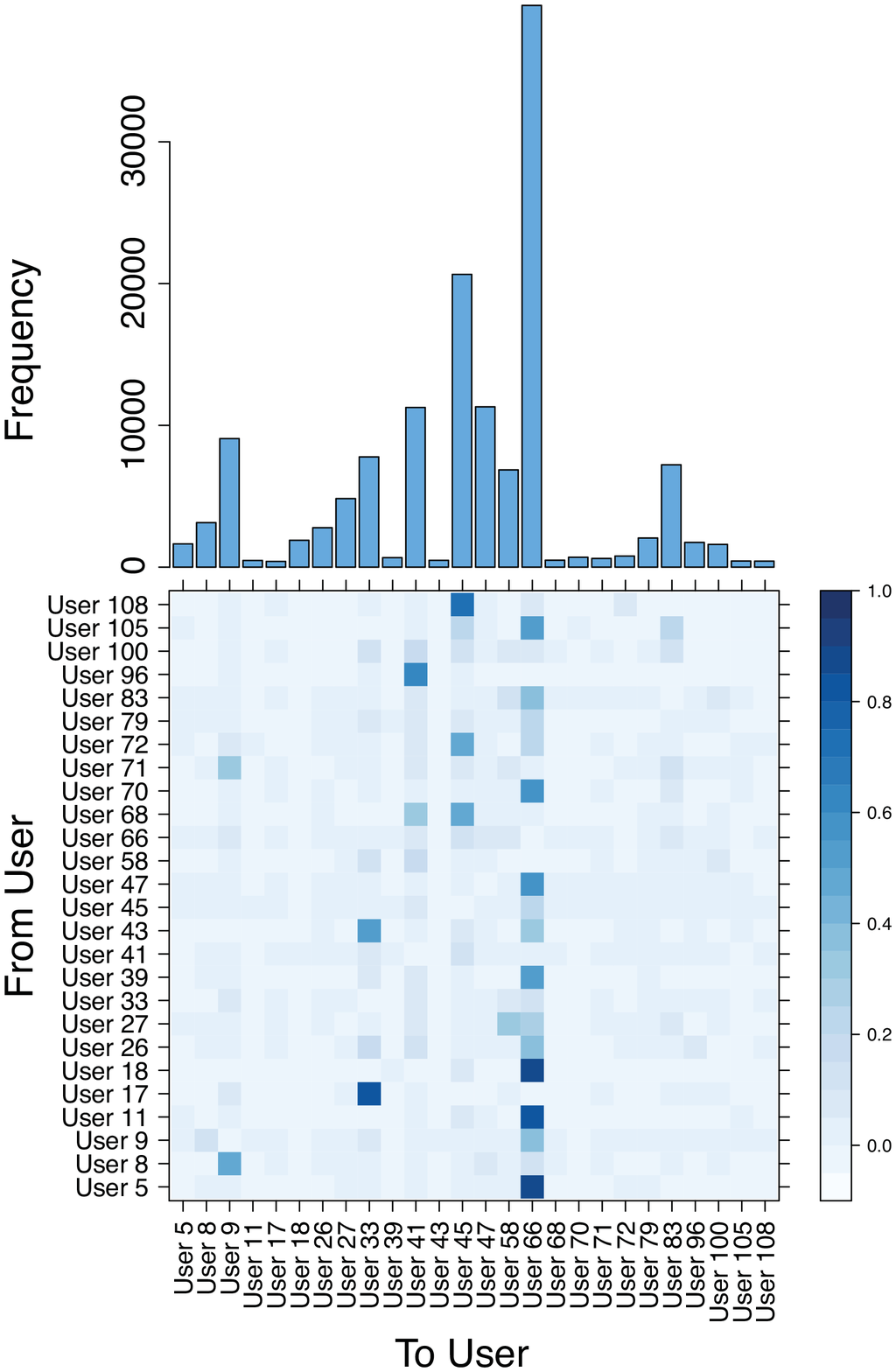}}
\subfigure[\captionictm]{\label{fig:ucsl:b}\includegraphics[width=\subfiguresizehoz\linewidth]{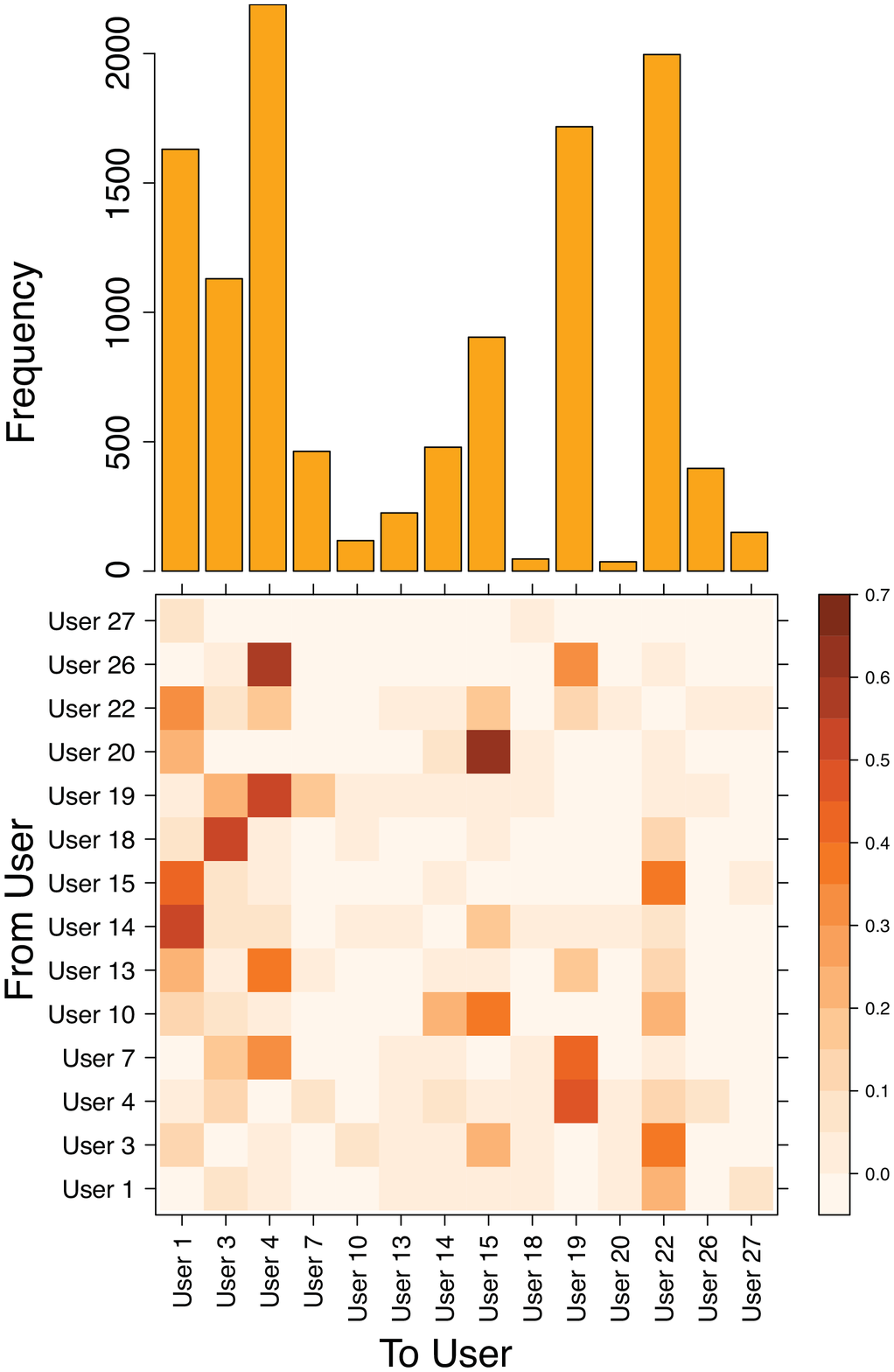}}
\subfigure[\captionncit]{\label{fig:ucsl:c}\includegraphics[width=\subfiguresizehoz\linewidth]{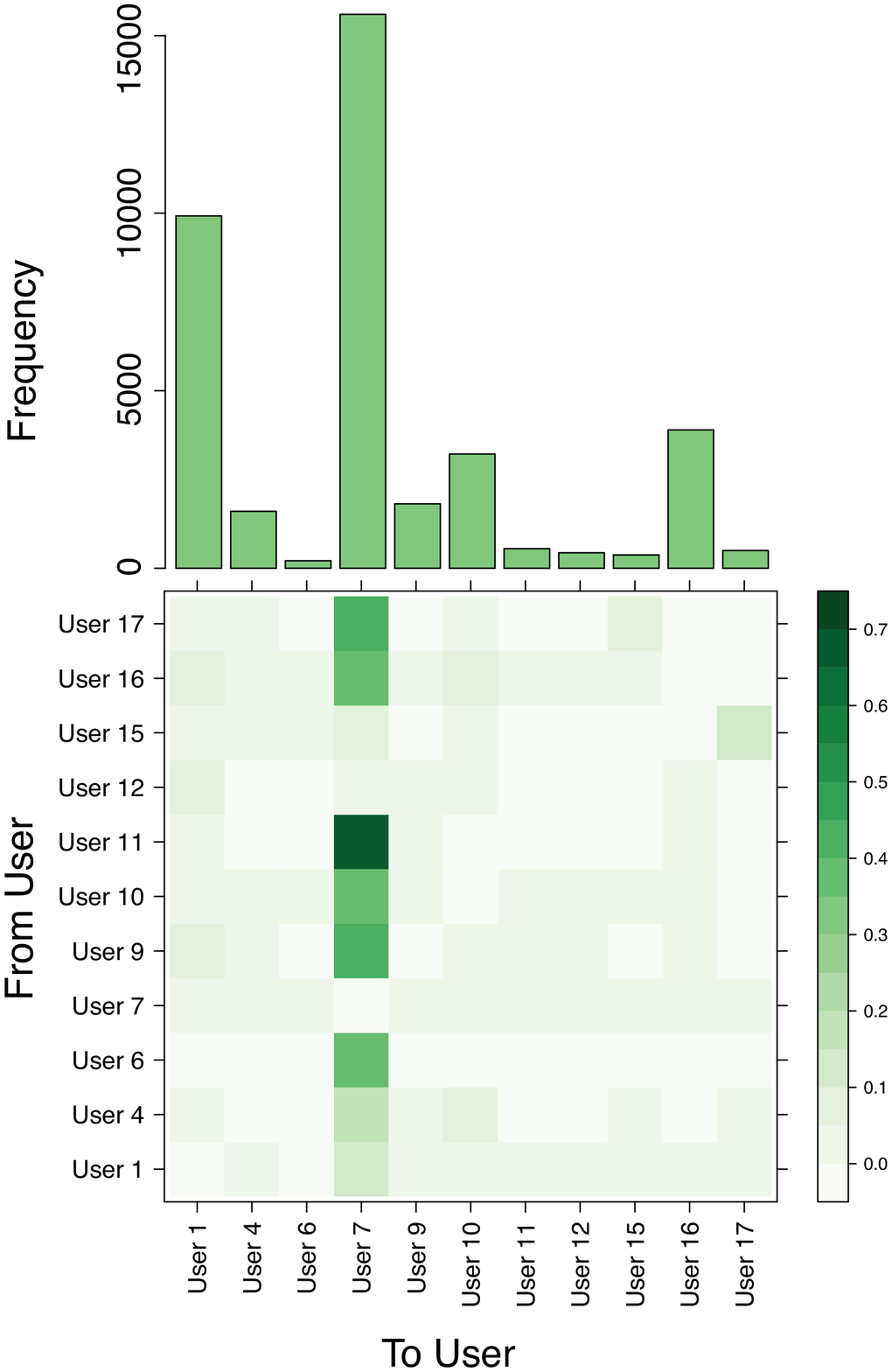}}\\
\subfigure[\captionbro]{\label{fig:ucsl:d}\includegraphics[width=\subfiguresizehoz\linewidth]{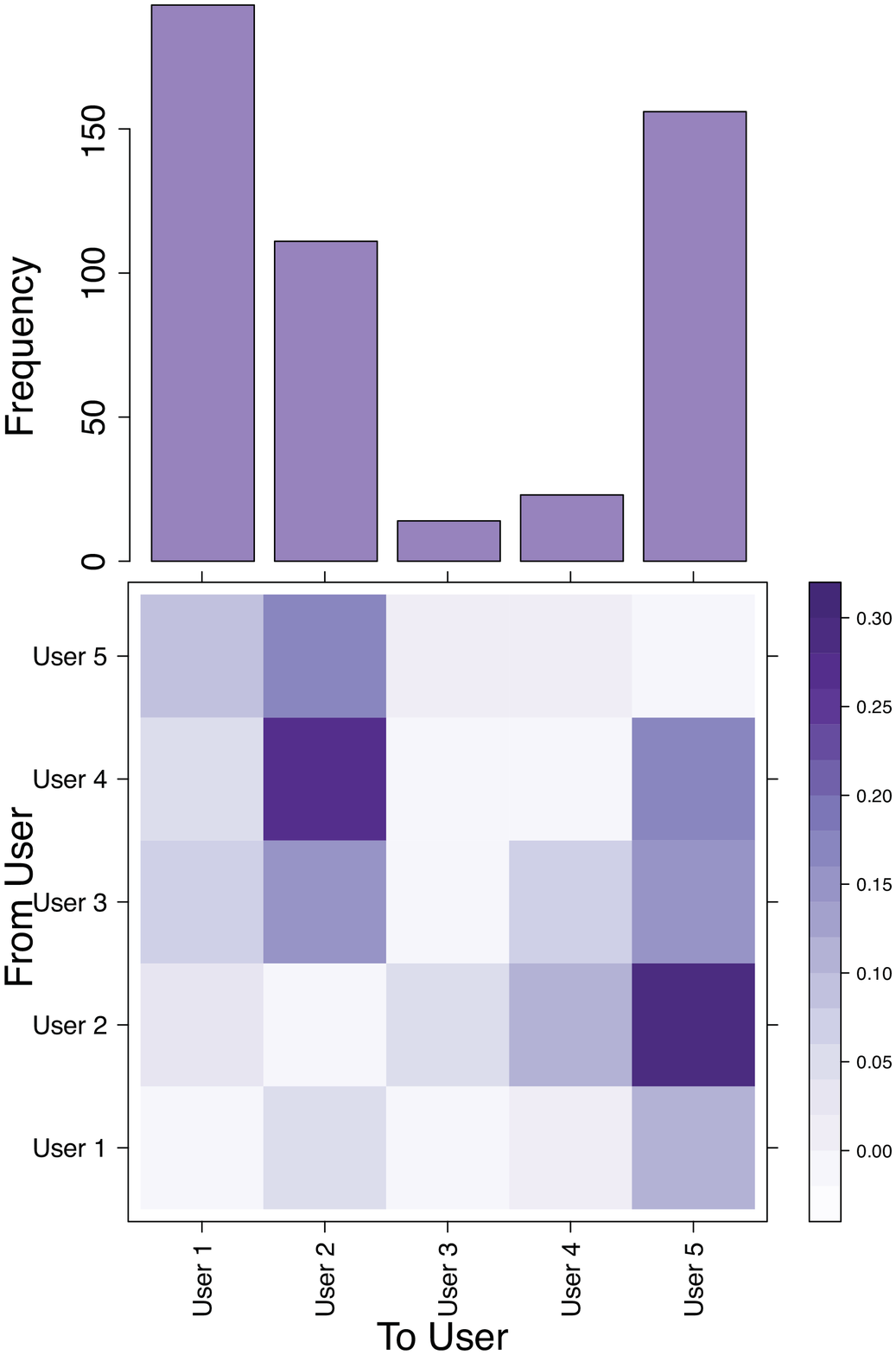}}
\subfigure[\captionopl]{\label{fig:ucsl:e}\includegraphics[width=\subfiguresizehoz\linewidth]{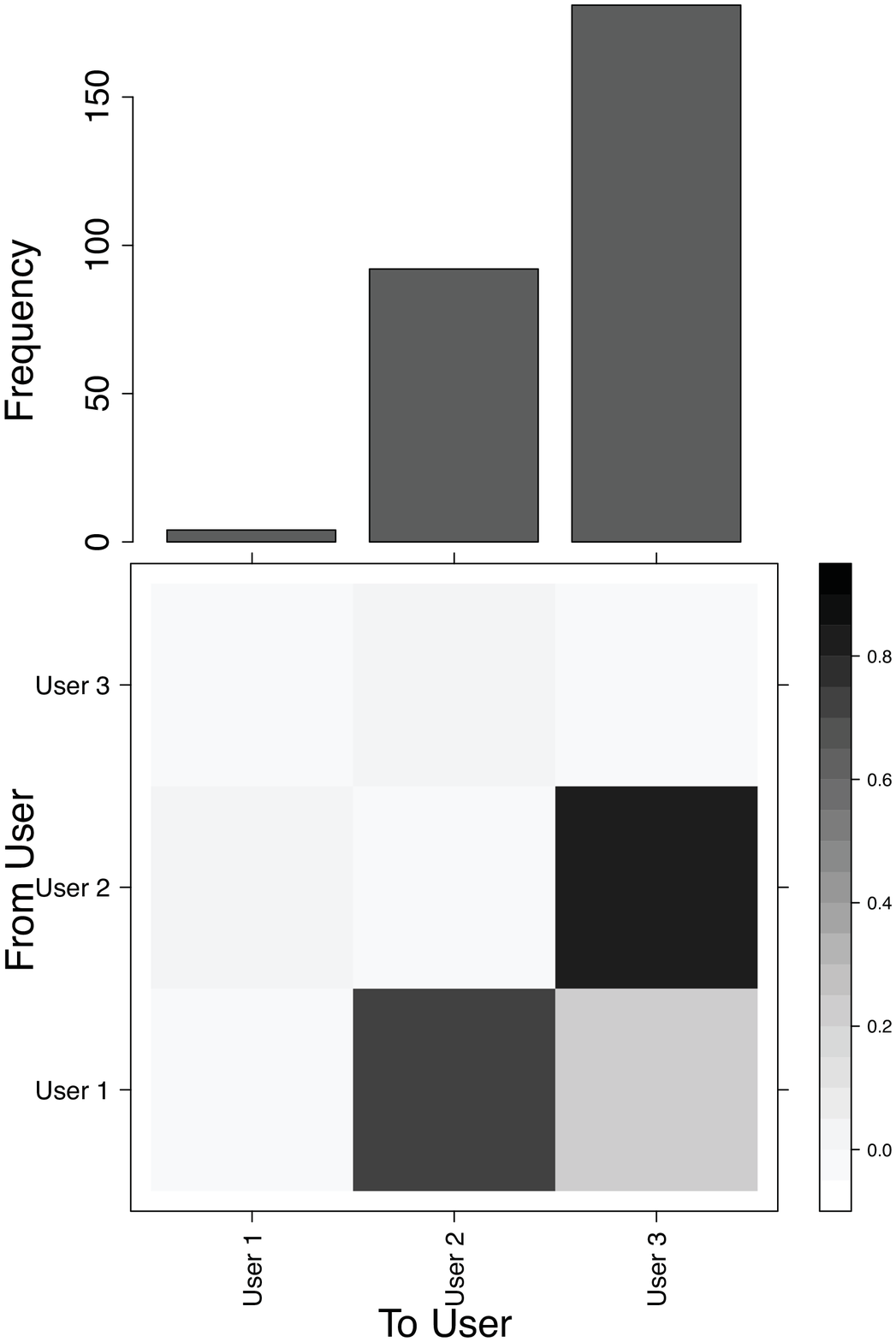}}
\caption{\textbf{Results for the \emph{User-Sequence Paths} analysis:} The columns and rows of the transition maps (\textbf{bottom area} of Figures~\ref{fig:ucsl:a} to \ref{fig:ucsl:e}) represent the transition-probabilities between the users of each dataset for a first-order Markov chain, where rows are \emph{source users} and columns are \emph{target users}. A sequence (or transition-probability) is always read \emph{from row to column}. Darker colors represent higher transition-probabilities while lighter colors indicate lesser transition-probabilities. Absolute probability values are dependent on the number of investigated rows and columns, hence relative differences are of greater importance.
Darker colored columns identify gardeners\green{, a contributor focused on pruning ontology classes and fixing syntactical errors.}
The histograms (\textbf{top area} of Figures~\ref{fig:ucsl:a} to \ref{fig:ucsl:e}) show the number of changes performed by each user (again for a first-order Markov chain) within the \numdatasets ontologies in alphabetical order. Note, that the $y$-axes for all histograms are scaled differently for each dataset. All datasets have a few users who contributed the majority of changes, while the rest of the users (the long-tail) only contributed a very small number of changes. Note that the transition-probabilities depicted in the transition maps are relative numbers for each column and row individually. \green{The sum of all transition probabilities for one row in the transition maps is 1. For example, if \emph{User 1} exhibits a transition probability of $0.30$ to another \emph{User 2} it means that \emph{User 2} has a $30\%$ probability of changing a class after \emph{User 1}.} Thus, an inspection of the transition maps \textbf{and} histograms is necessary for proper interpretation. To increase readability we have removed users from the plots who have contributed only a very limited number of changes for ICD-11, ICTM and NCIt.}
\label{fig:ucsl}
\end{figure*}

Analyzing the chronologically ordered list of contributors for each class of the \numdatasets investigated datasets provides the necessary information to identify users who perform changes on classes after (or before) other users. Note that this analysis on its own, without regarding additional factors, such as the changed property or the performed change-action, does not provide information about actual collaboration. The results of this analysis \green{could} be used to \green{potentially} identify users who work on the same classes, however, we do not know if they actually collaborate with or just clean up (i.e., a \emph{gardener}\green{, a contributor focused on pruning ontology classes and fixing syntactical errors}) after other users.

\green{
\textbf{Path \& model description:} To  analyze user sequences, we iterated over each class of our datasets and extracted a chronologically ordered list of contributors. For example, a given path for a given class can look like the following: \emph{User A, User B, User B, User C}. As we are interested in uncovering patterns of distinct users, we merged multiple consecutive changes by the same user into a single change -- our previous example would then unfold into: \emph{User A, User B, User C}. By doing so we remove biases emerging when one single user consecutively changes the same class over and over as this may result in unreasonable high transition probabilities between equal users.

We fit a first-order Markov chain model on this set of paths, where each path represents a single class of the ontology and each element of a path constitutes a change by a single user on the class. The resulting transition probabilities between users then e.g., tell us the probability that \emph{User B} changed a class after \emph{User A}.
Hence, they give us thorough insights into frequent consecutive user patterns that emerge when looking at which users contribute to classes in an ontology. Due to reasons of privacy we obfuscated the usernames and replaced them with generic names.
}

\textbf{Results:} When investigating the transition probabilities (representing a Markov chain of first order) between contributors (see bottom area of Figures~\ref{fig:ucsl:a} to \ref{fig:ucsl:e}) we can identify very active users by looking at darker colored columns of the transition maps. \green{Note that these darker colored columns can also be used to identify gardeners, a contributor focused on pruning ontology classes and fixing syntactical errors.} As we have merged all consecutive changes of the same user into one single change, the diagonal, representing the transition probabilities between the same users, is $0$. The absolute transition probabilities, depicted next to each transition map, are dependent on the absolute amount of observations and users, thus are to be interpreted relatively to each other for each row individually. When looking at the probabilities between the \green{three most active users (being users 66, 45 and 47)}, and all corresponding target users in ICD-11 we can see that the probabilities are very evenly distributed among them. \green{Meaning that, when investigating the rows (\emph{From User}) that correspond to the top three most active users, probabilities to all target users (\emph{To User}) are very evenly distributed, with very minor exceptions. This indicates that users who contribute many changes to ICD-11 are not followed by specific other contributors, but exhibit an even distribution of users that edited a class after them.}
Nonetheless, we can clearly identify \emph{User 66} to be the most likely user that edits a class after nearly all other users. This suggests, that \emph{User 66} may  represent a gardener\green{, a contributor focused on pruning ontology classes and fixing syntactical errors}, in ICD-11. 

For NCIt we can clearly observe that \emph{User 7} appears to be a \emph{gardener}, who is checking all the changes contributed by all other users. For BRO \emph{Users 2} and \emph{5} are prominent target users, evident in the high transition probabilities as \emph{To User} (dark columns) -- i.e., they frequently edit a class after other users do. Interestingly, the user with the highest number of changes (\emph{User 1}) exhibits very low and evenly distributed transition probabilities (row) and is not necessarily the user that most likely changes a class after another users. This shows us that there does not need to be a necessary connection between the overall activity of users and their activity as a gardener. \green{This could also mean that \emph{User 1} is possibly working independently from the other users in BRO, or that \emph{User 1} is a domain specialist and all other users only change concepts that have not been worked on by that specialist. However, further investigations in future work are required to confirm this observation as our Markov chain analysis is not able to determine this kind of distinction.} For OPL we can observe that \emph{User 3} frequently changes the same classes after \emph{User 2}. A similar observation can be made for \emph{Users 1} and \emph{2}. However, one has to keep in mind that \emph{User 1} has contributed a limited number of changes, rendering the observed transition probabilities less useful as they rarely occur.

\green{
The histograms (see top area of Figures~\ref{fig:ucsl:a} to \ref{fig:ucsl:e}) indicate that  a small number of users contribute the majority of changes (similar to a long-tail distribution). 
However, this appears to be more dominant for specific ontologies compared to others. 
In order to measure the inequality among contributions of changes to a specific ontology by users, we analyzed the \emph{Normalized Entropy}\footnote{Additionally, we calculated the \emph{Gini Coefficient} for each distribution confirming the results presented here.}, which is determined by calculating the \emph{Shannon Entropy} and normalizing the entropy by dividing by the logarithm of the length (i.e., number of users) of a distribution. This coefficient measures the statistical dispersion of a distribution -- i.e., the coefficient is one if all users contributed equally to the ontology, while it is zero in case of total inequality where a single user conducts all changes. The results indicate that ICD-11 ($0.55$) exhibits a low entropy value, i.e., the changes are dominated by only a few users. For NCIt ($0.61$), OPL ($0.64$) and ICTM ($0.68$) we receive medium normalized entropies indicating a more democratic contribution to the ontology by users. A high entropy can be observed for BRO ($0.81$), which indicates that it is a demographically edited ontology -- even though there are only five users.\footnote{Note that we do not necessarily know whether the differences between these distributions are statistically significant as we are mainly interested in the behavior of single distributions.} 
}

\textbf{Interpretation \& practical implications:}
The transition probabilities for a first-order Markov chain unveil the roles of certain users and can help to identify users or even groups of users who frequently change the same classes. Users that frequently change classes after other users (i.e., exhibit high transition probabilities in their columns) were identified by us as actual gardeners, curators and administrators of the corresponding projects. If certain users always change the same classes after specific other users, it could be worthwhile for project administrators to investigate if these users are actually collaborating, for example by looking at the changed properties and property values, or if a single user is always cleaning up after the other user. In all datasets we were able to observe at least one user who contributed a high number of changes, with evenly distributed transition probabilities to all remaining users.
This observation indicates that in all projects, gardeners, curators and administrators are assigned (directly or indirectly) certain parts of the ontology; otherwise the transition probabilities between the very active users would be higher.

The ability of understanding who is most likely going to change a specific class next, as well as the classes that a user is most likely to change next \green{could be} used by project administrators to help users in finding and identifying classes (and thus work) of interest. On the other hand, the information about the next, most probable contributor for a class, can even be used to create automatic class recommender systems to suggest work to users, which could help to increase participation. \green{However, these two analyses are beyond the scope of this paper and are therefore subject to future work.} In particular for projects the size of ICD-11 and NCIt, mechanisms to automatically identify and assign work are highly useful as it is still very time-consuming to find pending work and users with the necessary knowledge to address the identified work-tasks.

\subsection{Structural paths}
\label{sub:structural paths}

The investigation of \emph{Structural Paths} involves an analysis of different aspects regarding how and where users contribute to the ontology, such as the depth level of the class that users contribute to next (Section~\ref{subsub:depth-level paths}) as well as looking at the relationship distances between consecutively changed classes (Section~\ref{subsub:hierarchical-relationship paths}).

\subsubsection{Depth-level paths}
\label{subsub:depth-level paths}

In this analysis, we investigate if users concentrate their efforts on specific depth levels of the ontology and if there are certain depth levels that are frequently consecutively changed and receive less concentrated workflows. The gathered results provide the necessary information to implement prefetching mechanisms, \green{potentially} helping to minimize the loading and waiting times for contributors. Furthermore, we can determine whether users move along the structure of the underlying ontology when editing classes.

\begin{figure*}[!ht]
\centering
\subfigure[\captionicd]{\label{fig:dluc:a}\includegraphics[width=\subfiguresizehoz\textwidth]{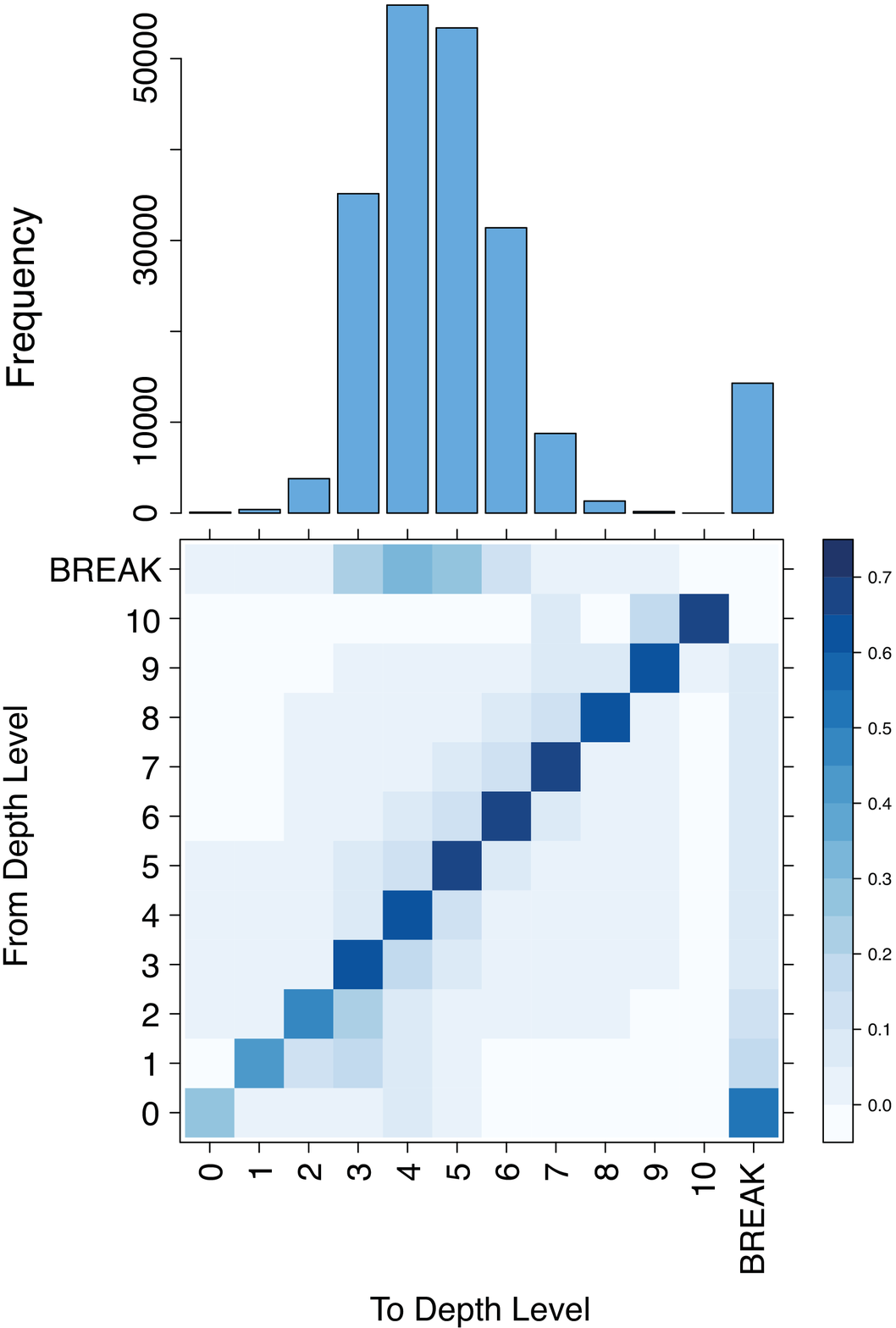}}
\subfigure[\captionictm]{\label{fig:dluc:b}\includegraphics[width=\subfiguresizehoz\textwidth]{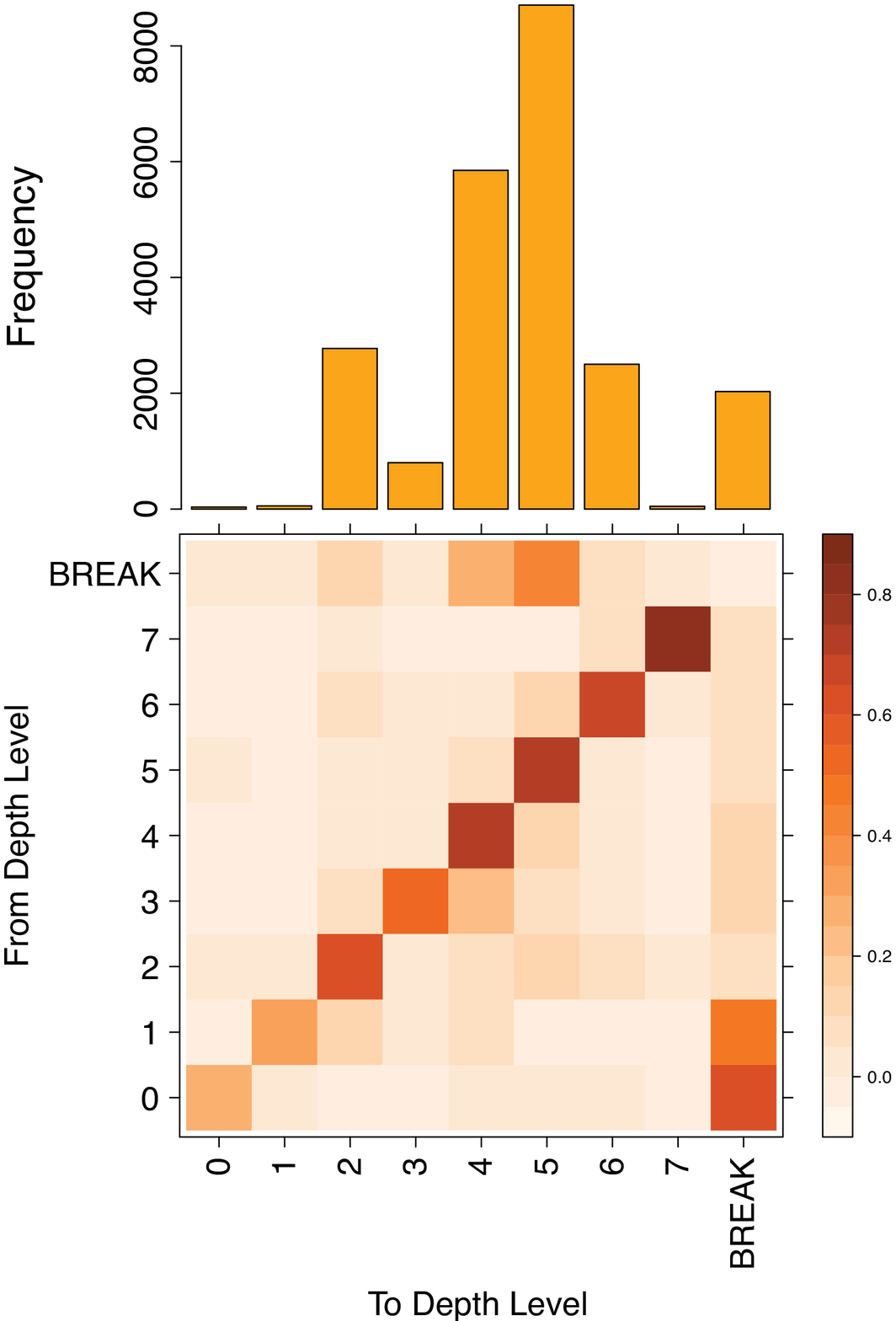}}
\subfigure[\captionncit]{\label{fig:dluc:c}\includegraphics[width=\subfiguresizehoz\textwidth]{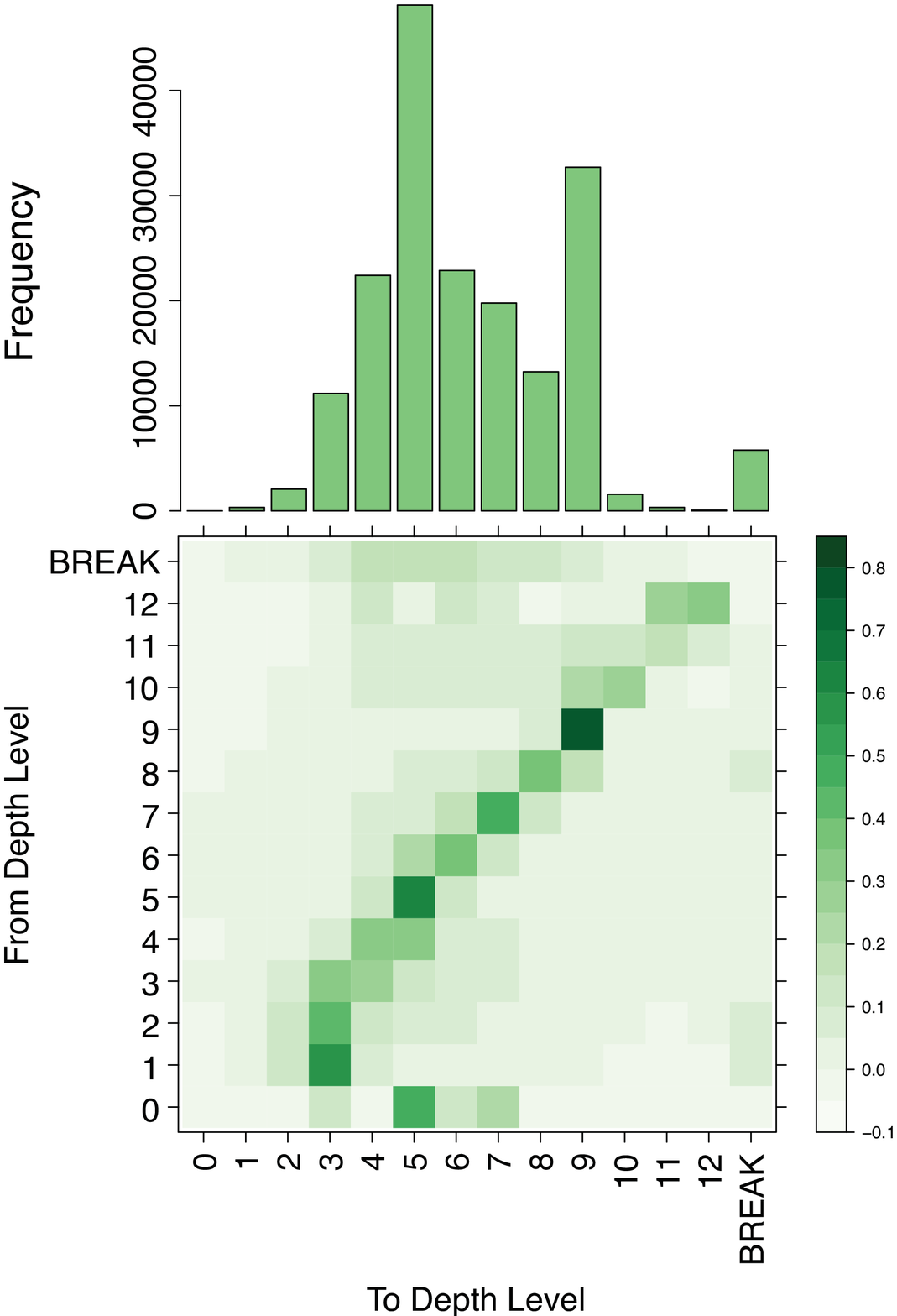}}\\
\subfigure[\captionbro]{\label{fig:dluc:d}\includegraphics[width=\subfiguresizehoz\textwidth]{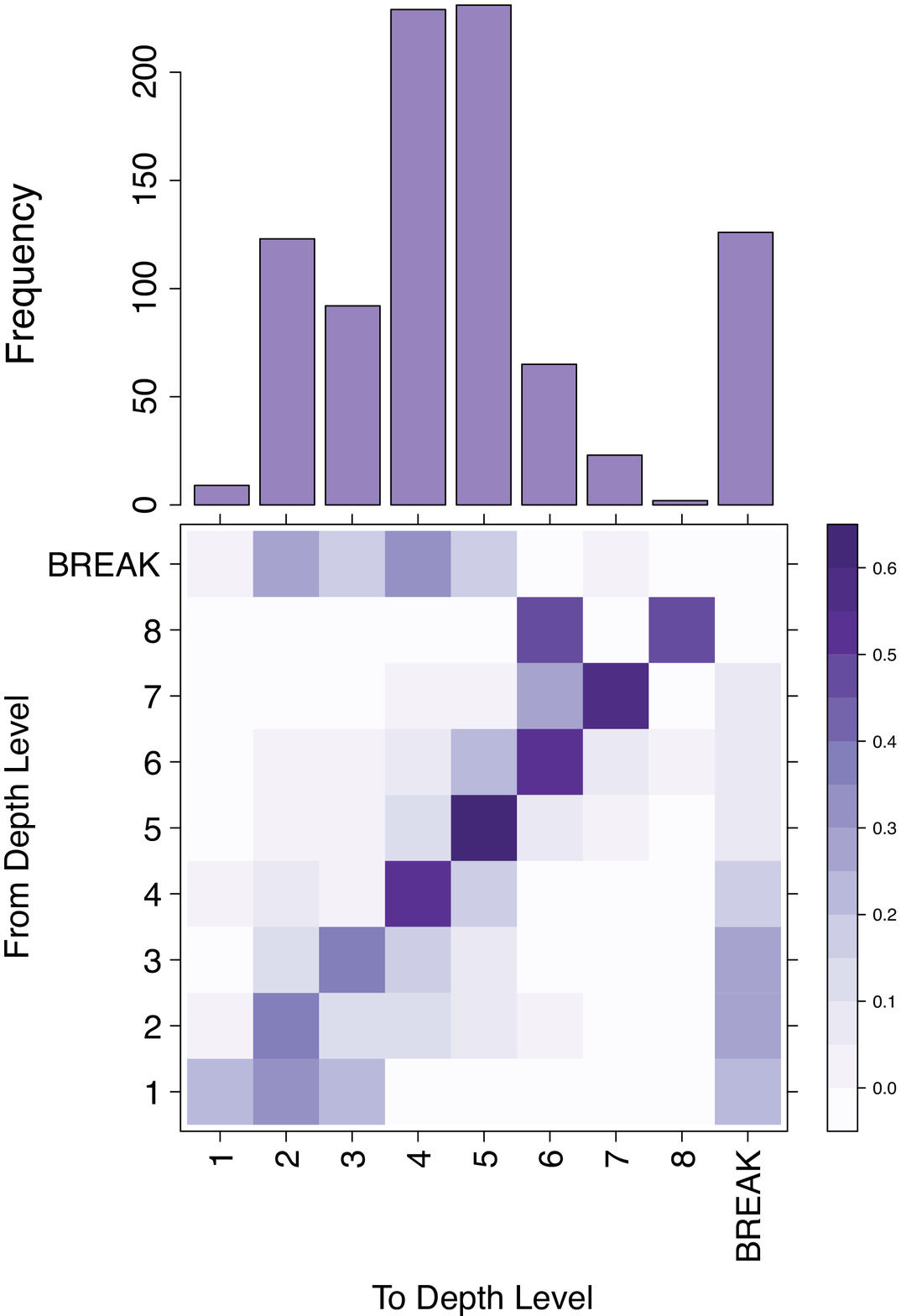}}
\subfigure[\captionopl]{\label{fig:dluc:e}\includegraphics[width=\subfiguresizehoz\textwidth]{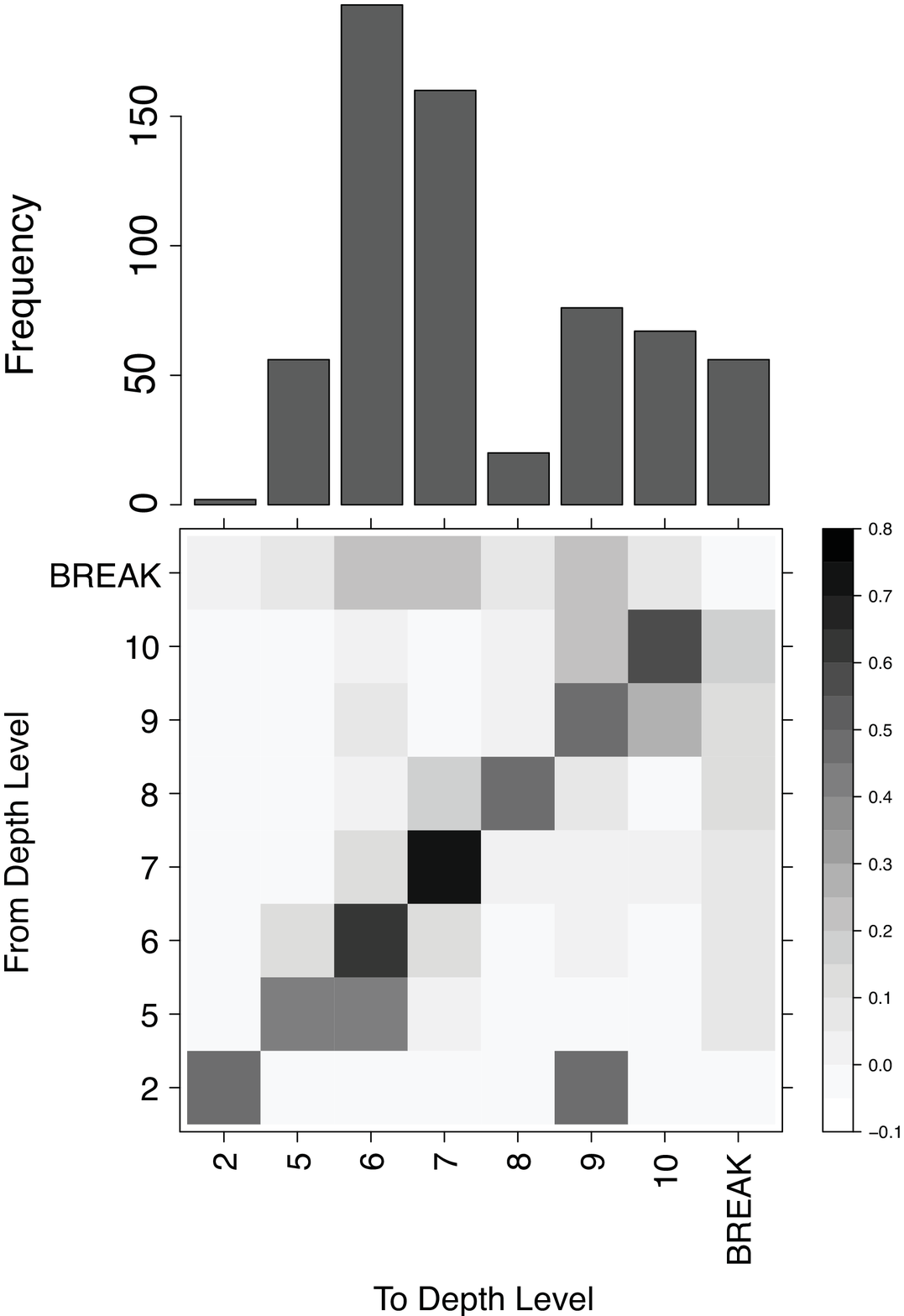}}
\caption{\textbf{Results for the \emph{Depth-Level Paths} analysis:} The columns and rows of the transition maps (\textbf{bottom area} of Figures~\ref{fig:dluc:a} to \ref{fig:dluc:e}) represent the transition probabilities of a first-order Markov chain between depth levels, where rows are \emph{source depth levels} and columns are \emph{target depth levels}.
A sequence (or transition probability) is always read \emph{from row to column}. Darker colors represent higher transition probabilities while lighter colors indicate lesser transition-probabilities. Absolute probability values are dependent on the number of investigated rows and columns, hence relative differences are of greater importance.
For classes closer to root a \emph{top-down} editing manner can be observed, while this is reversed for classes further away from root. \green{The sum of all transition probabilities for one row in the transition maps is 1. For example, if \emph{Depth-Level 6} exhibits a transition probability of $0.30$ to another \emph{Depth-Level 5} it means that a class on \emph{Depth-Level 5} has a $30\%$ probability of being changed after a class on \emph{Depth-Level 6}.}
The histograms (\textbf{top area} of Figures~\ref{fig:dluc:a} to \ref{fig:dluc:e}) show the number of changes performed in each depth level aggregated over all users of the respective projects (again for a first-order Markov chain). Throughout all projects, classes located between the first and last few depth levels (in the middle) are changed substantially more frequently than others, suggesting that work is concentrated on some depth levels while others receive none to very few changes at all.
Note, that the $y$-axes for all histograms are scaled differently for each dataset. For the $x$-axes (and column/rows of the transition maps) we only display depth levels which exhibit at least one change, thus, the depth level sequences are not necessarily continuous from lowest to highest depth level.}
\label{fig:dluc}
\end{figure*}

\begin{figure*}[!ht]
\centering
\setcounter{subfigure}{5}
\subfigure[\captionicd]{\label{fig:dluc:f}\includegraphics[width=\subfiguresizehoz\textwidth]{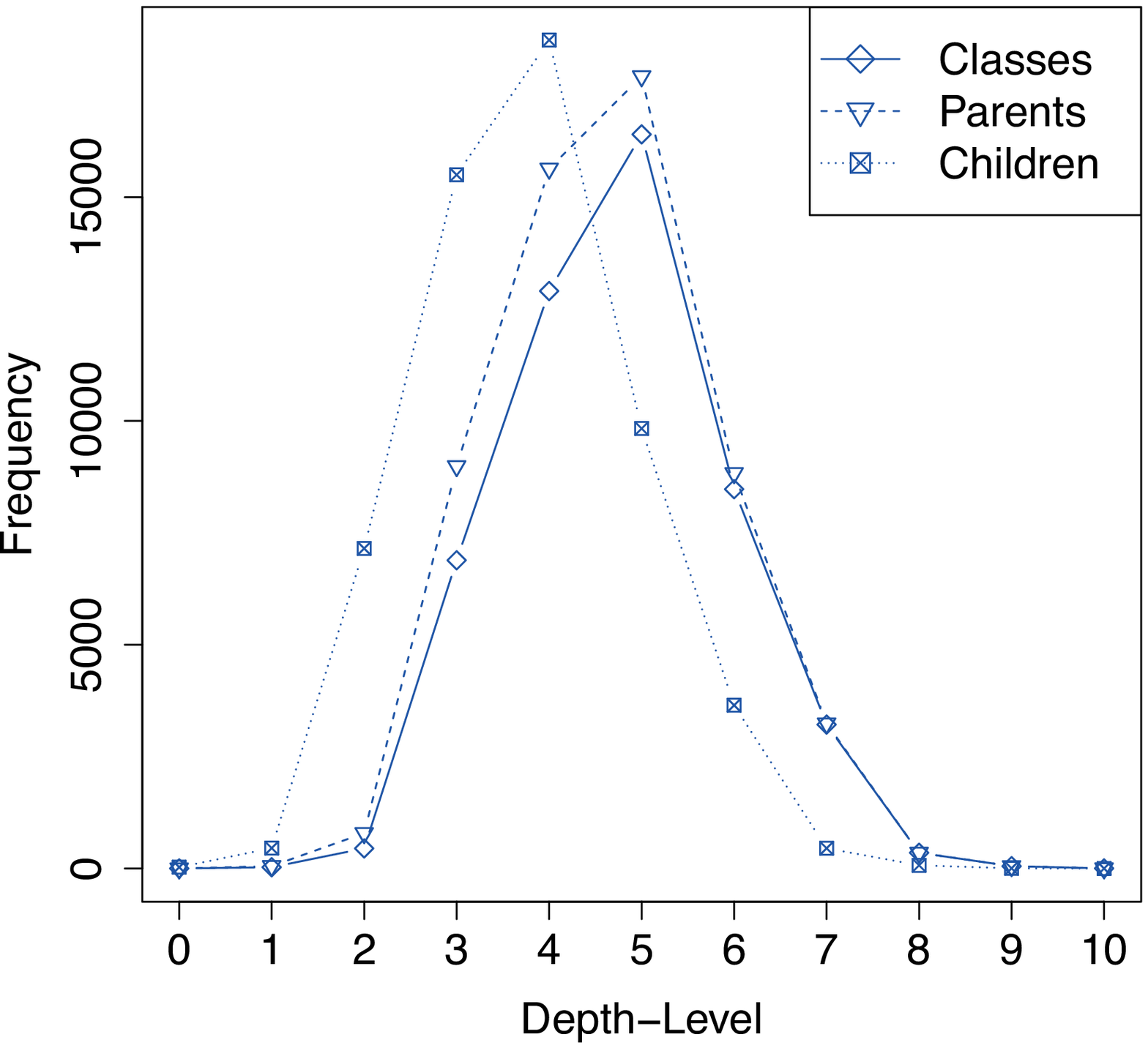}}
\subfigure[\captionictm]{\label{fig:dluc:g}\includegraphics[width=\subfiguresizehoz\textwidth]{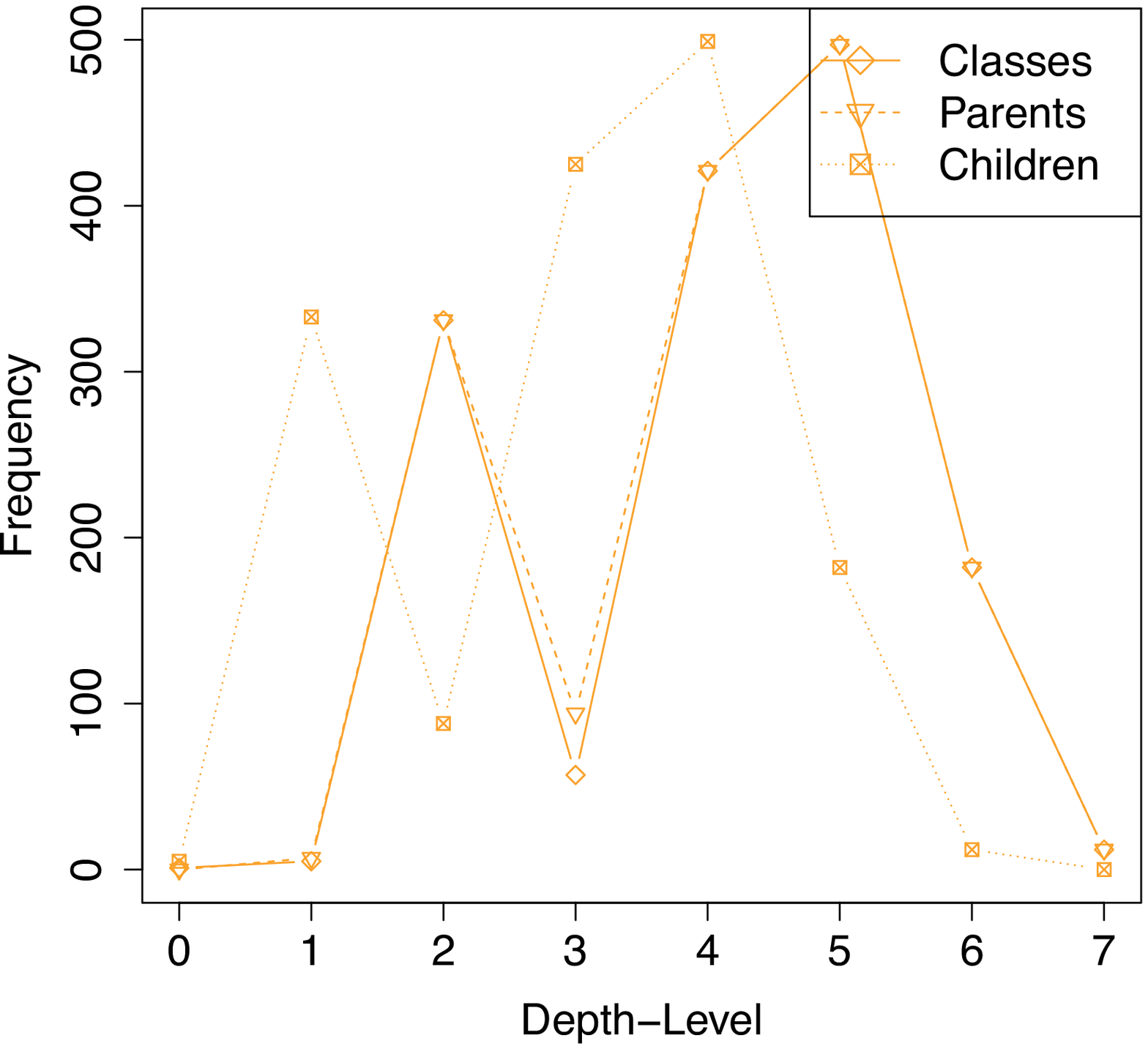}}
\subfigure[\captionncit]{\label{fig:dluc:h}\includegraphics[width=\subfiguresizehoz\textwidth]{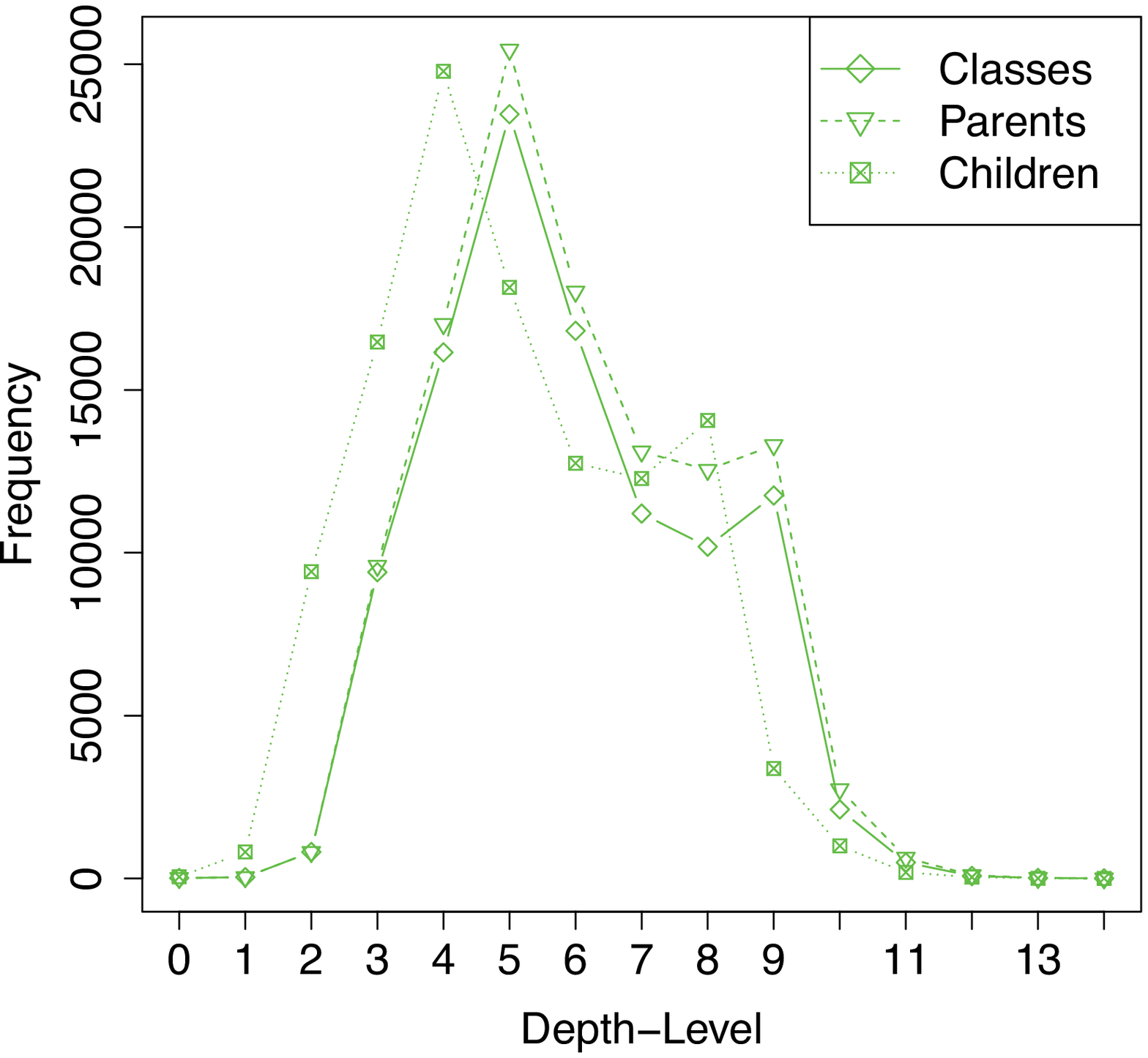}}\\
\subfigure[\captionbro]{\label{fig:dluc:i}\includegraphics[width=\subfiguresizehoz\textwidth]{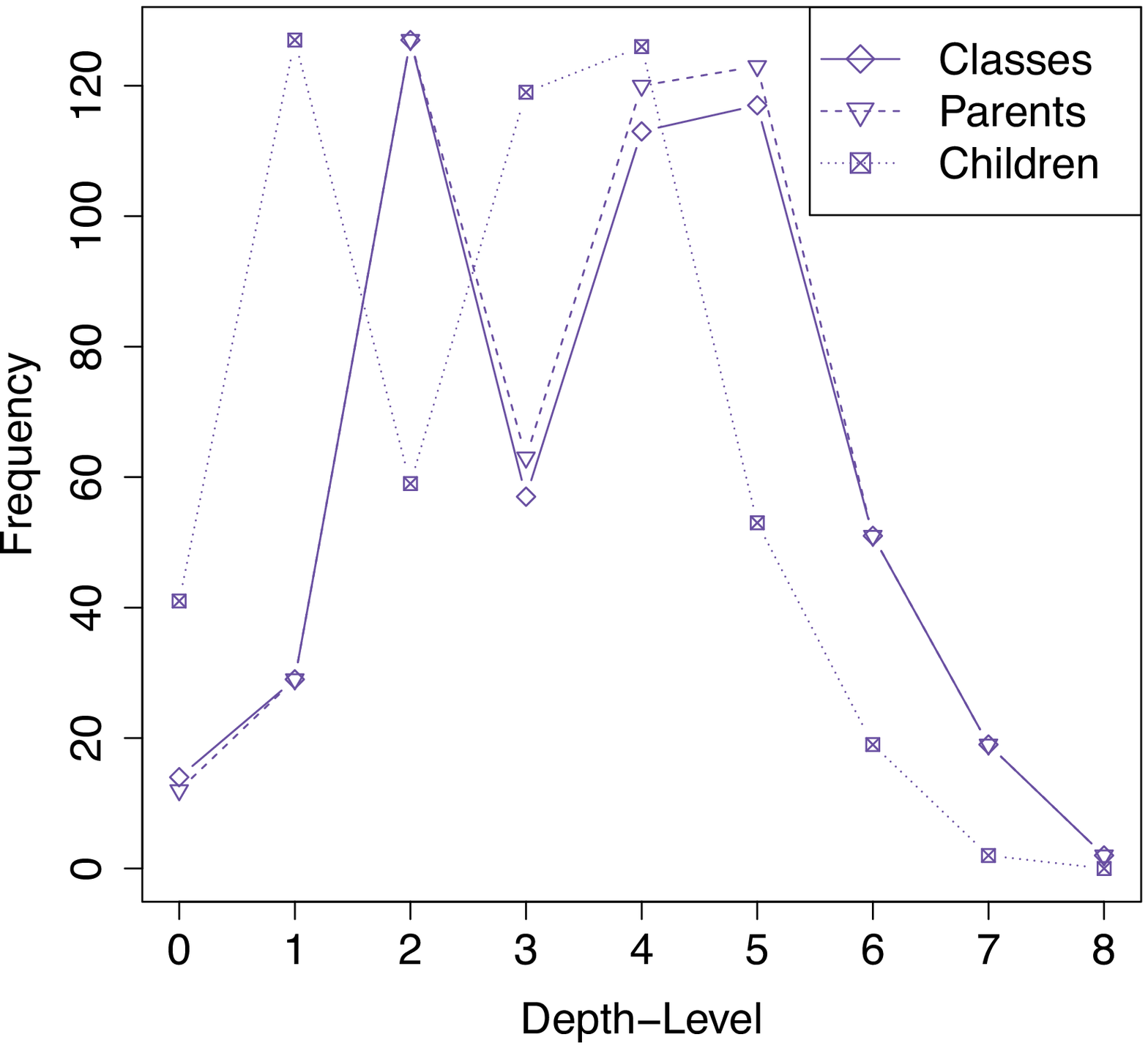}}
\subfigure[\captionopl]{\label{fig:dluc:j}\includegraphics[width=\subfiguresizehoz\textwidth]{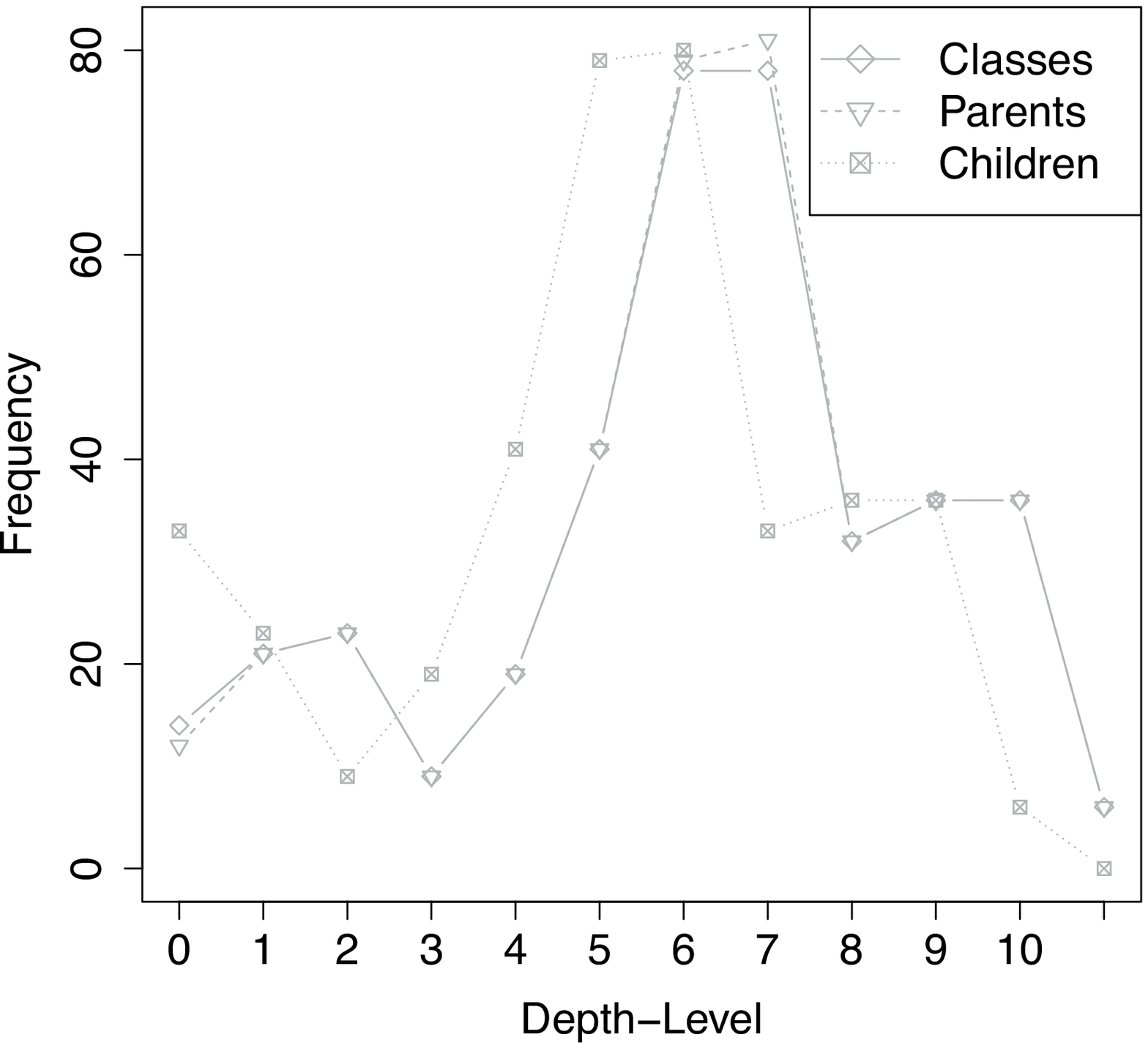}}
\caption{The \textbf{Figures~\ref{fig:dluc:f} to \ref{fig:dluc:j}} depict the absolute numbers ($y$-axis; Frequency) of classes as well as the number of edges (\emph{isKindOf}) to classes on the immediate higher (\emph{parents}; closer to root) and lower (\emph{children}; further away from root) depth level for all depth levels ($x$-axis; Depth-Level). According to Figures~\ref{fig:dluc:f} to \ref{fig:dluc:j} the transition probabilities depicted in the transition maps correlate with the total number of edges to children and parents for each depth level across all datasets.}
\label{fig:dluc:res}
\end{figure*}

\green{
\textbf{Path \& model description:} For this analysis, we stored  the chronologically ordered depth levels of each changed class for each user (user-based). The depth level  of a class is  the length of the shortest path  between the \emph{root node} of the ontology and the corresponding class. For example, a given path for a given user can look like the following: \emph{Depth 3 (for class A), Depth 3 (for class A), Depth 3 (for class A), Depth 3 (for class B), Depth 4 (for class C)}. We merged consecutive changes that were conducted by the same user on the same class into one single sequent change between the same depth levels. Hence, for our previous example we would merge the three successive changes of class A into just two consecutive ones which results in the following final depth-level path: \emph{Depth 3, Depth 3, Depth 3, Depth 4}. This approach helps us to investigate patterns of changing distinct depth levels while still retaining the notion of users consecutively editing the same classes.

Consequently, we fit a first-order Markov chain model on these paths -- each path represents a single user and each element of a path represents a corresponding depth level of a class the user has changed. The final transition probabilities give us information about consecutive depth levels that users change over time. For example, they might tell us the probability that users change a class belonging to the third depth level of the ontology after one that has a depth level of $2$.
}

\textbf{Results:} First, the histograms (see top area of Figures~\ref{fig:dluc:a} to \ref{fig:dluc:e}) show that work is concentrated on certain depth levels of the ontology, with the highest and lowest levels not receiving as much attention as the levels in-between.

\green{As depicted in the transition maps (bottom area of Figures~\ref{fig:dluc:a} to \ref{fig:dluc:e}), users have a high tendency to edit classes in the same depth levels, visible in the darker colored diagonal. In ICD-11, for the first five depth levels, users appear to have a tendency towards \emph{top-down} editing, \green{evident in the darker immediately right of the diagonal,} while this tendency \green{turns around into a \emph{bottom-up} editing behavior, evident in the darker colored squares immediately left of the diagonal,} at a depth level of $6$ and higher, and appears to be strictly limited to surrounding depth levels.} For ICTM (see Figure~\ref{fig:dluc:b}), we can observe a similar trend, again with the tendency towards \emph{top-down} editing appearing to be minimally more dominant.
For NCIt, when only looking at the transition map, we can identify a trend towards \emph{bottom-up} editing\green{, evident in the squares directly left of the diagonal being darker than the ones right of the diagonal.} \green{However, when also considering
the absolute number of changes, depicted in the histogram of Figure~\ref{fig:dluc:c}, we can infer that the levels with a higher frequency of occurrence, even though their transition probabilities are more evenly distributed, have a greater impact on the editing strategy. This means that while we can see a \emph{bottom-up} editing behavior for levels $8$ to $5$ and a \emph{top-down} editing behavior for levels $1$ to $4$, classes on levels $1$ to $4$ are more frequently changed than classes on the other levels, hence a tendency towards \emph{top-down} editing can be observed.} Thus, when users are not changing the same classes, they still exhibit a preference towards \emph{top-down} editing. Given the short observation periods for BRO and OPL it is hard to infer edit strategies. However, similar to the other projects, we can observe a concentration on the same depth levels with alternating preferences towards higher and lower depth levels. Similar to ICD-11, all datasets exhibit higher transition probabilities between the immediately surrounding depth levels. 

Furthermore, we investigate whether the total number of classes as well as the total number of links to the immediate higher (children; edges to classes one level further away from root) and lower (parents; edges to classes one level closer to root) depth level correlate with our findings (Figures~\ref{fig:dluc:f} to \ref{fig:dluc:j}). For example, the transition map for ICD-11 (see Figure~\ref{fig:dluc:a}) shows that contributors exhibit a \emph{top-down} editing behavior for the first five depth levels, with level $5$ exhibiting first signs of \emph{bottom-up} editing. Figure~\ref{fig:dluc:f} shows a higher number of possible transitions from children than parents, indicating that users are in general likelier to follow \emph{top-down} editing-strategies when changing classes, following relationships by chance, of the first four levels. This changes for ICD-11 at level 5, with a higher number of transitions to parents than to children, and continues until level 10. Resulting in a higher probability of users performing \emph{bottom-up} editing-strategies when changing classes from levels 6 to 10. The same observations can be made for all other datasets, indicating that the class hierarchy influences the edit behavior of contributors.

In all datasets, after taking a \emph{BREAK} (representing an artificially introduced session break when two consecutive changes of the same user are more than $5$ minutes apart; for more information see Section~\ref{sub:limitations}), users exhibit a clear tendency towards changing classes on certain depth levels (e.g., levels 3 to 5 for ICD-11, levels 4 to 5 for ICTM,  levels 4 to 7 for NCIt, levels 2 to 4 for BRO and levels 6 to 9 for OPL).

\textbf{Interpretation \& practical implications:} \green{The results of this analysis show if, to what extent and where (limited to locality being determined by \emph{isKindOf} relationships) work is conducted and concentrated within the ontology. This information can \green{potentially} be used in a variety of ways, for example by ontology-engineering tool developers to adapt the interface of the ontology-engineering tool dynamically to display specific classes after users return from a \emph{BREAK}.
Project managers can adapt milestones and project progress reports to reflect the underlying editing strategies (e.g., \emph{top-down} editing), for example by aligning progress with created branches (opposed to complete coverage). 
Another potential use-case for the results of this analysis involves the prefetching of content in certain environments (e.g., mobile or embedded systems) to minimize waiting times.}
Across all projects we can observe that classes close to and very far away from the \emph{root} of the ontology are not edited as frequently as other classes. One explanation for this observation could be that classes in lower depth levels (closer to \emph{root}) are mainly used as content dividers and are usually created in the beginning of a project. Thus, they may be more stable and less frequently updated. Classes at the higher depth levels (further away from \emph{root}) on the other hand most likely require extensive expert knowledge. Hence, only a small number of users have the necessary expertise to contribute to these classes. Additionally, the absolute number of classes in the higher and lower depth levels is much lower in all investigated datasets. Note that absolute values of depth levels are less important for the interpretation of the results than their relative position (i.e., closest to root, furthest away from root, etc.). For example, a class at level $6$ can exhibit different behaviors in ontologies with $6$ or $10$ levels.

In all projects, except for NCIt, the depth levels where users start to edit the ontology after they return from a \emph{BREAK} are similar to the ones where they stop editing before taking a \emph{BREAK}. To be able to make that observation we have to take the absolute numbers of changes on each depth level (bottom area of Figure~\ref{fig:dluc}) into account when looking at the transition probabilities (top area of Figure~\ref{fig:dluc}). NCIt is the only dataset where users appear to be similarly likely to take a \emph{BREAK} after changing classes across all depth levels, except for $0$ and $12$.

When we combine the results of this analysis with the results of the \emph{User-Sequence Paths} (Section~\ref{sub:activity paths}) we may be able to develop automatic mechanisms to curate and delegate work to users. For example, if we know that a specific user is most probably going to contribute to a class on level $3$ and we have a set of classes on that level where that specific user is the most probable next user to contribute to, determined by the \emph{User-Sequence Paths} analysis, we may combine these two observations to create class (and thus work) suggestions for users.

\subsubsection{Hierarchical relationship paths}
\label{subsub:hierarchical-relationship paths}

Given the high number of observed transitions between the same depth levels in the \emph{Depth-Level Paths} analyses (Section~\ref{subsub:depth-level paths}; bottom area of Figure~\ref{fig:dluc}), we conducted an additional analysis investigating the relationships between the changed classes for all users. \green{Hence, we wanted to know if all worked-on classes on the same depth-levels are siblings, cousins or any other kind of close relative? And in general, can we determine if users follow these hierarchical orders of an ontology when contributing to classes on the same depth level?}
\green{To further strengthen our observation that users are actually moving along the ontological hierarchy when contributing to an ontology (see Section~\ref{subsub:depth-level paths}), we analyzed the relationships between the changed classes for each user. Note that whenever we talk about relationships for this analysis, we refer to the hierarchical \emph{isKindOf} relationships between two classes, e.g., parent, child, sibling or cousin.} For example, when traversing the shortest-path distance of $2$, multiple different nodes can be reached, such as a grandparent (i.e., 2 times up), a grandchild (i.e., 2 times down), a sibling (i.e., 1 time up, 1 time down) or even some other relationship (e.g., 1 time down, 1 time up).

\green{
\textbf{Path \& model description:} By combining the information from the \emph{Depth-Level Paths} and the relative movement between depth levels, we inferred the hierarchical relationships between two consecutively changed classes of a single user (user-based). For example, if the difference between the depth levels of the investigated classes would be exactly the size of the shortest-path between them (with the shortest-path being $>0$), the latter-changed class could either be a \emph{Child}, a \emph{Parent}, an \emph{Ancestor} or a \emph{Descendent} of the first-changed class. Given a relative \emph{DOWN} movement (to a lower depth level) value, depending on the shortest-path value, the second class could be classified as \emph{Child} (shortest-path of $1$) or \emph{Descendent} (shortest-path $>1$). Analogously follows the definition of a \emph{Parent} and \emph{Ancestor} with a relative \emph{UP} movement. A \emph{Sibling} is defined as the two classes being (i) connected via the same parent with (ii) a shortest-path distance of $2$ and (iii) both classes are located on the \emph{SAME} depth level.
A \emph{Cousin} is used when two classes on the \emph{SAME} depth level are connected by the same grand parent while exhibiting a shortest-path distance of $4$. Every other possible combination of depth level and shortest-path was classified as \emph{Other}.
\emph{Self} indicates that the same class that was changed last time was changed again. For example, a consecutive change of \emph{Sibling} and \emph{Self} means that a change was first performed on a class that is a sibling of the previous class (not displayed in this example) and then another change was performed on the same class, however now the relationship changed to \emph{Self} as no new class was involved.

Again, consecutive changes on the same class by the same user have been merged into one single sequent change (c.f. Section~\ref{subsub:depth-level paths}), meaning that multiple (more than $2$) consecutive changes of the same user on the same class have been merged into \emph{Self} to \emph{Self}.
Hence, a given path for a single user can, e.g., look like the following: Sibling, Self, Self, Child.

We fit a first-order Markov chain model to the data -- each path represents a single user and each element represents a hierarchical relationship between the classes changed by the user. The resulting transition probabilities of the fitted model can then give us insights into common emerging patterns. E.g., we can identify how probable it is that users change a \emph{Sibling} after a \emph{Child}.
}

\begin{figure*}[!ht]
\centering
\subfigure[\captionicd]{\label{fig:rsuc:a}\includegraphics[width=\subfiguresizehoz\textwidth]{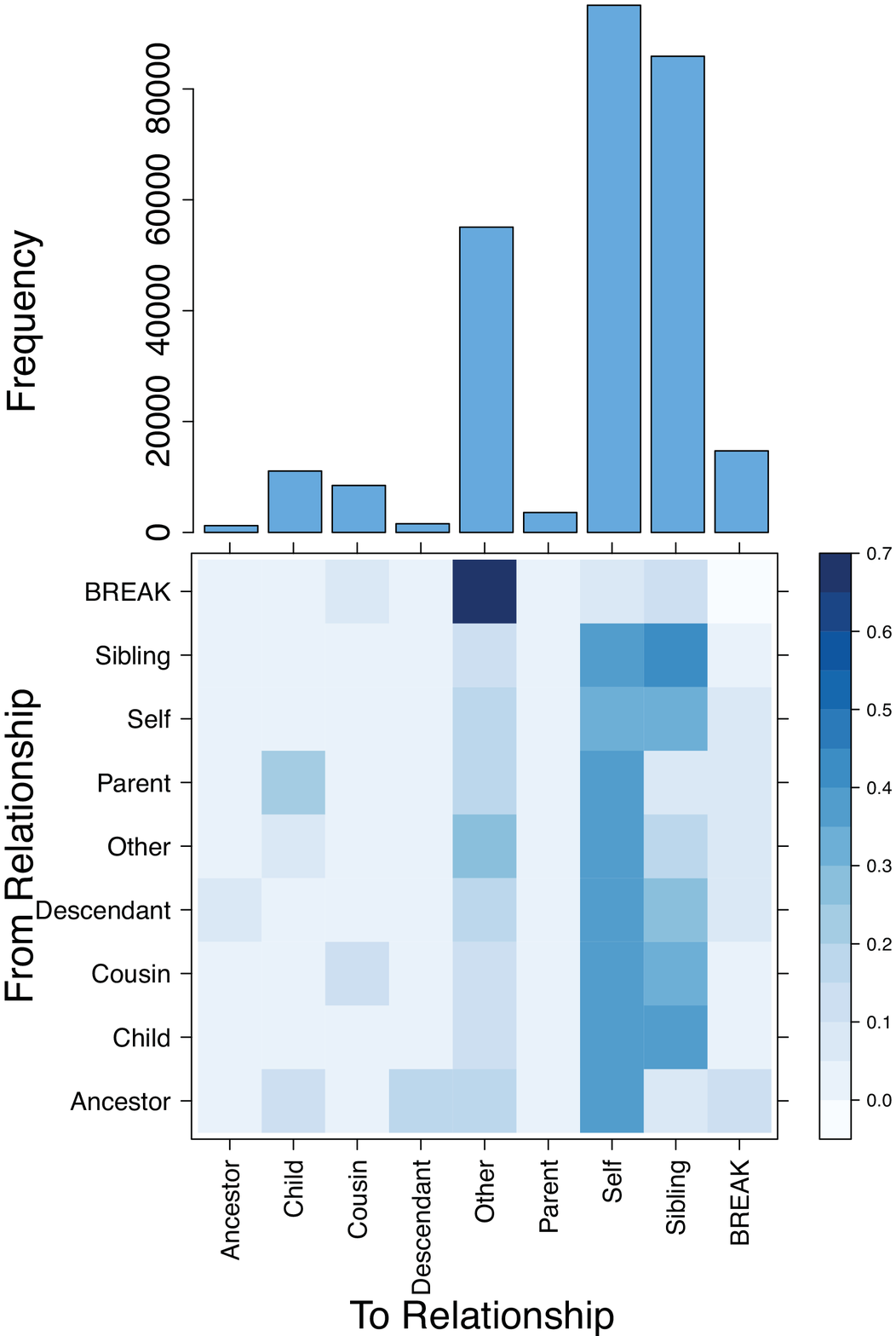}}
\subfigure[\captionictm]{\label{fig:rsuc:b}\includegraphics[width=\subfiguresizehoz\textwidth]{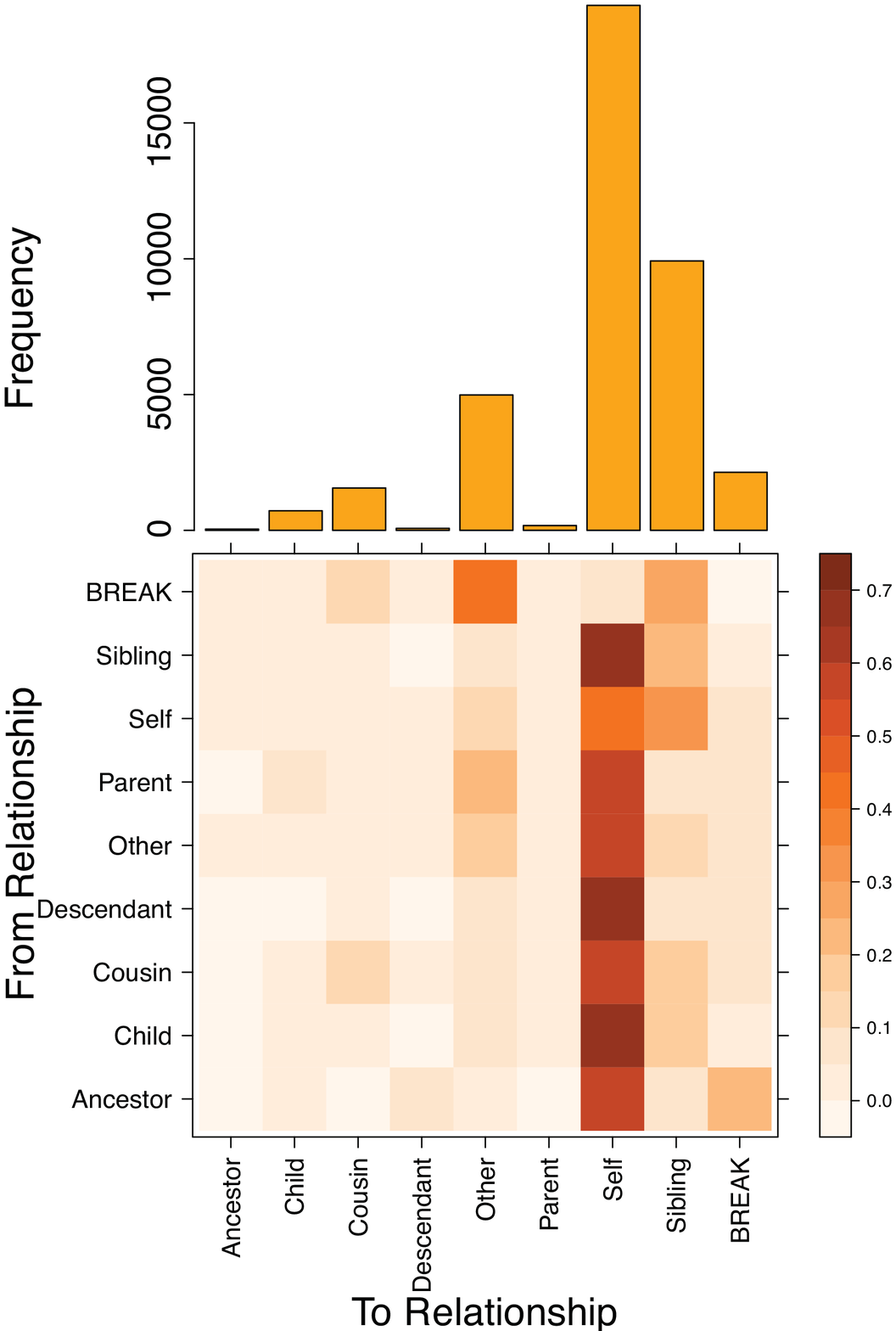}}
\subfigure[\captionncit]{\label{fig:rsuc:c}\includegraphics[width=\subfiguresizehoz\textwidth]{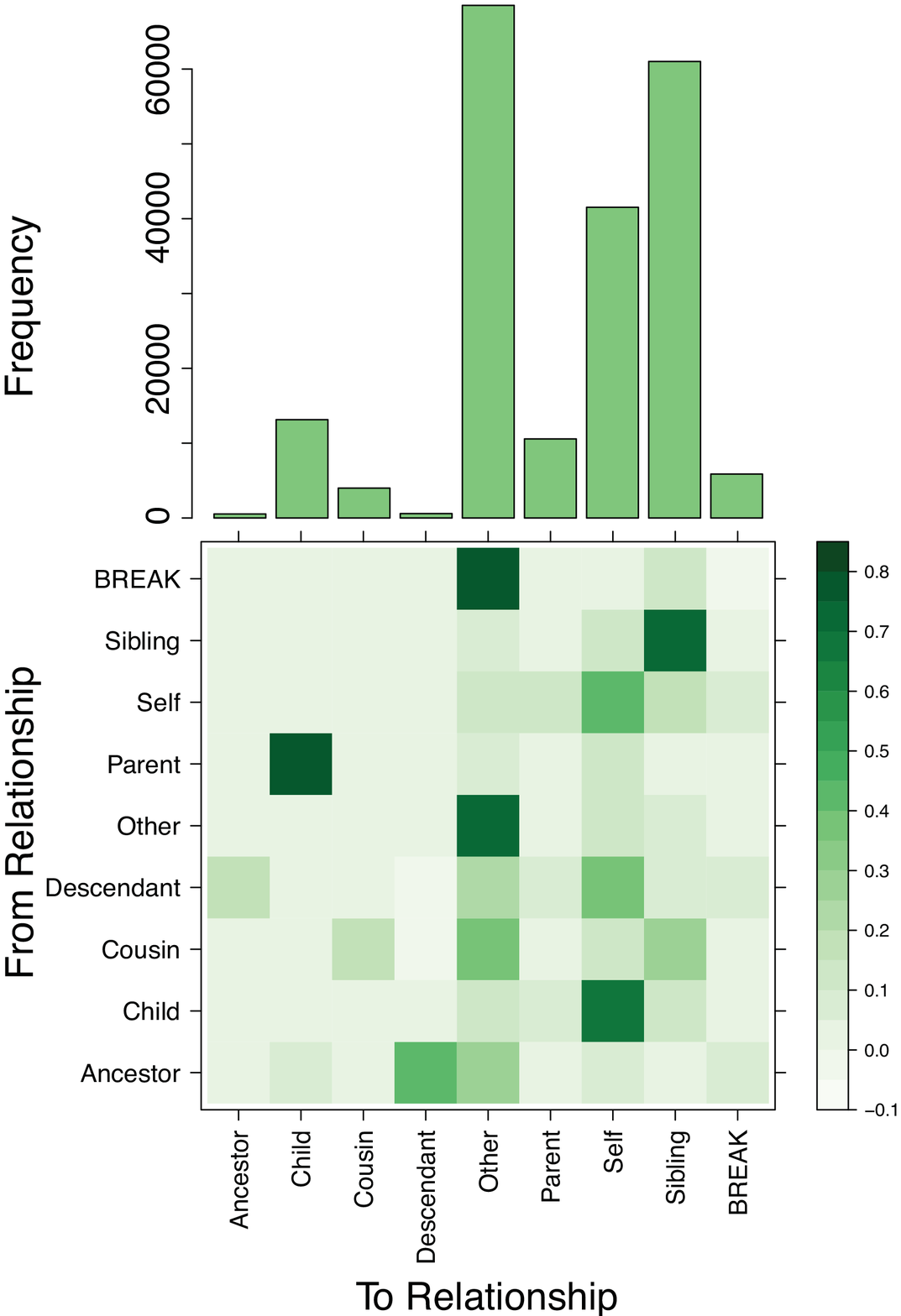}}
\subfigure[\captionbro]{\label{fig:rsuc:d}\includegraphics[width=\subfiguresizehoz\textwidth]{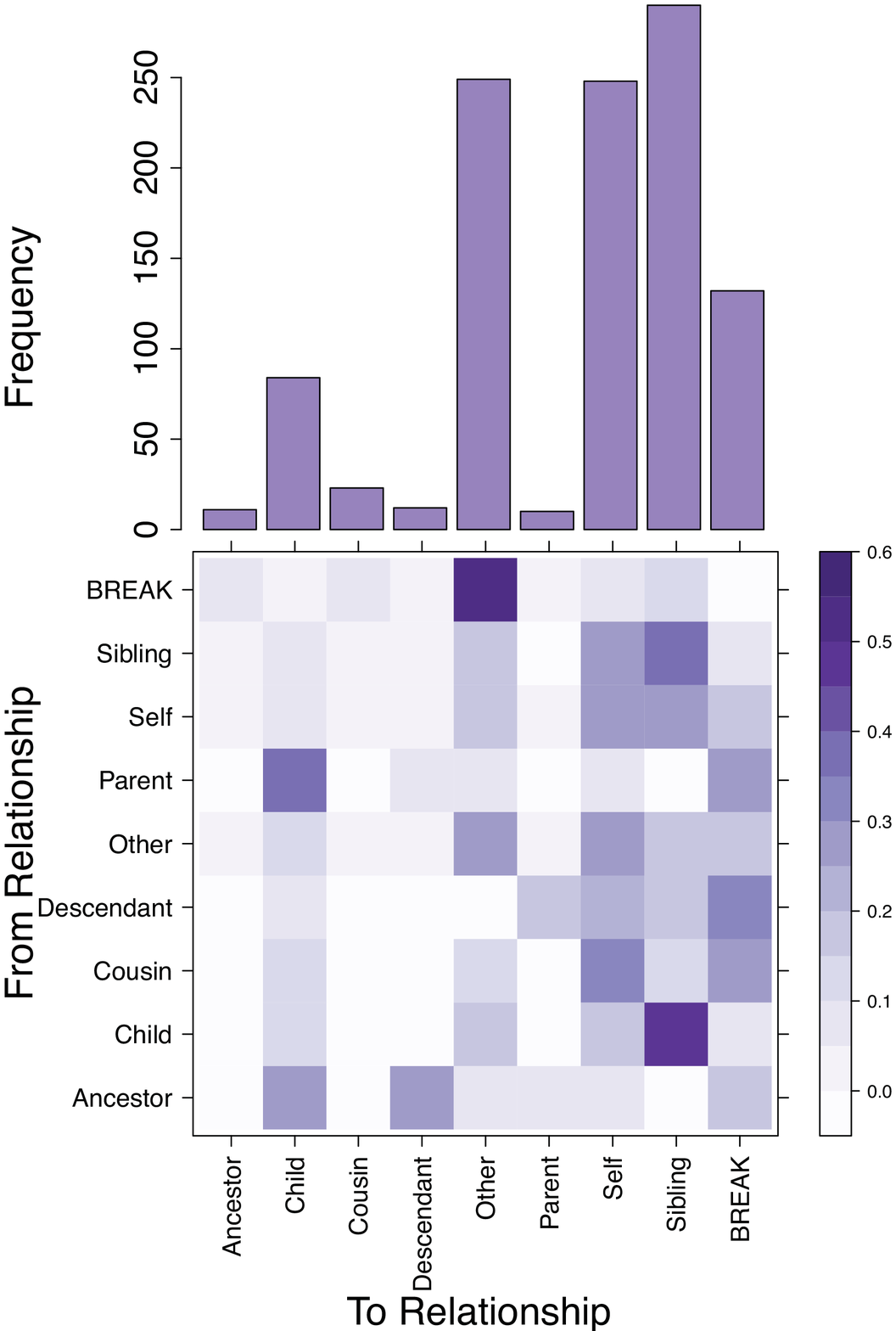}}
\subfigure[\captionopl]{\label{fig:rsuc:e}\includegraphics[width=\subfiguresizehoz\textwidth]{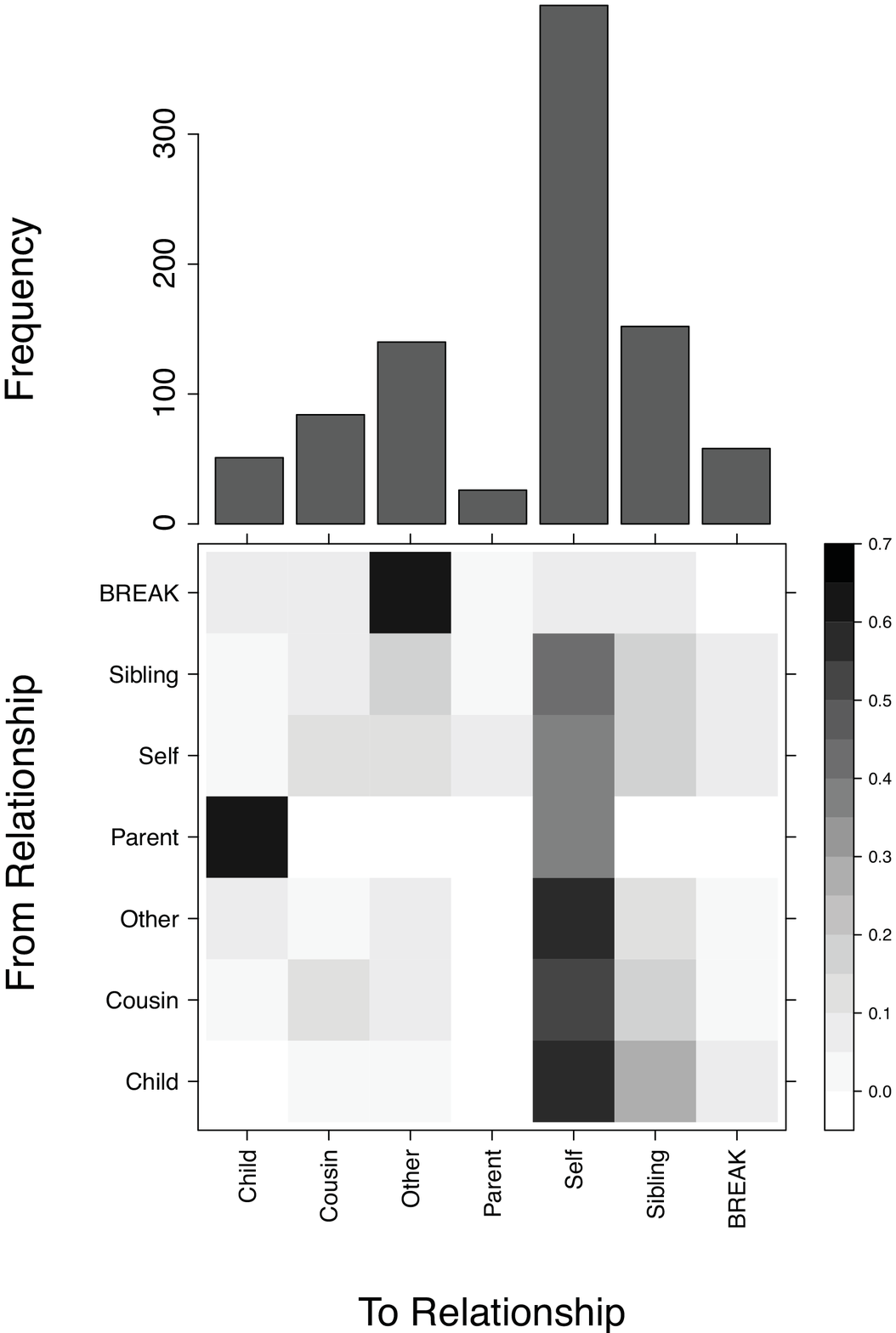}}
\caption{\textbf{Results for the \emph{Hierarchical-Relationship Paths} analysis:} The columns and rows of the transition maps (\textbf{bottom area} of Figures~\ref{fig:rsuc:a} to \ref{fig:rsuc:e}) represent the transition-probabilities of a first-order Markov chain between hierarchical-relationship levels, where rows are \emph{source relationships} and columns are \emph{target relationships}.
A sequence (or transition-probability) is always read \emph{from row to column}. Darker colors represent higher transition-probabilities while lighter colors indicate lesser transition-probabilities. Absolute probability values are dependent on the number of investigated rows and columns, hence relative differences are of greater importance. Across all datasets, aside from \emph{Self}, a very clear trend towards editing the ontology along \emph{Siblings} can be observed.
The histograms (\textbf{top area} of Figures~\ref{fig:rsuc:a} to \ref{fig:rsuc:e}) show the total number of occurrences of each relationship in the corresponding datasets aggregated over all users (again for a first-order Markov chain).
Note, that the $y$-axes for all histograms are scaled differently for each dataset. For the $x$-axes (and column/rows of the transition maps) we only relationships that occur at least once in the corresponding paths, thus the $x$-axes could be different from project to project. Given the very high amount of \emph{Self} and \emph{Sibling} transitions we can concur that users, when they contribute to classes on the same depth level follow a \emph{breadth-first} strategy\green{, meaning that they first concentrate their work on closely related classes (\emph{Siblings}) on the same depth-level before switching to a different branch on the same or any other depth-level.}}
\label{fig:rsuc}
\end{figure*}

\textbf{Results:} When looking at the histograms (see top area of Figures~\ref{fig:rsuc:a} to \ref{fig:rsuc:e}), we can observe that the relationships \emph{Self}, \emph{Sibling} and \emph{Other} are highly represented across all datasets.
The transition maps (bottom area of Figures~\ref{fig:rsuc:a} to \ref{fig:rsuc:e}) show that after a \emph{BREAK}, across all \numdatasets datasets, users tend to change classes ``somewhere els'' in the ontology, evident in the high transition probability from \emph{BREAK} towards \emph{Other}, and are likely not to  resume work in the same area of the ontology that they stopped working on. For ICD-11, ICTM and OPL, no matter which relationship type occurs, users tend to  edit the same class consecutively (dark colors in the \emph{Self} column). From this \emph{Self} relationship, which is also the one that occurs the most often in ICD-11, ICTM and OPL, users are very likely either to  change the same class again (\emph{Self}) or to change a \emph{Sibling} of the current class.

For NCIt, BRO and OPL we can observe that users, when changing a \emph{Parent} are very likely to change a \emph{Child} of that parent afterwards. Note, that this \emph{Child} does not necessarily have to be the same class that was changed prior to the traversal to \emph{Parent}. In all datasets, except for OPL, very high transition probabilities towards \emph{Other} can be observed for all not so frequently present relationships. In particular for NCIt we can observe that \emph{Other} is the most frequently observed transition, even before \emph{Self} and \emph{Sibling}.

\textbf{Interpretation \& practical implications:} By combining the results of this analysis with the results of the \emph{Depth-Level Paths} analysis, we can infer that users exhibit a tendency towards \emph{top-down} editing while contributing to the ontology, when only considering changes that occur on different depth levels. If they concentrate their efforts on the same depth levels, users exhibit a \emph{breadth-first} editing behavior\green{, meaning that they first concentrate their work on closely related classes (\emph{Siblings}) on the same depth-level before switching to a different branch on the same or any other depth-level}, either changing the same class multiple times or traversing along siblings of the current class. We can leverage this information  not only  to refine the previously suggested pre-fetching of classes but also to enhance possible class recommendations. Similarly, it is possible for ontology-engineering tool developers to minimize the necessary efforts of users to contribute to the ontology by implementing, for example, guided workflows that take the underlying edit strategies of the contributors into account.

As classes in ICD-11 and ICTM have a large number of properties and for ICTM certain properties have to be added in multiple languages, the high transition probabilities towards \emph{Self} (dark colors in the \emph{Self} column) are not surprising.
One possible explanation for this observation for ICD-11 could be the special functionality available in iCAT (for ICD-11) that allows users to export parts of the ontology as spreadsheets for local editing and adding property values. Once contributors finished editing the spreadsheet they have to  enter the data into the system manually, as no automatic import functionality is present. In the iCAT interface, users are simultaneously presented with the ontology tree for navigating through the classes and the corresponding properties and property values. When users select a property they can easily switch between classes, with the selected property staying selected, thus allowing to quickly enter the same properties for different classes.

A similar, yet not as dominant as in ICD-11 and ICTM, behavior can be observed for NCIt and BRO and even to some extent in OPL, which all do not use the export functionality. According to our observations, users travel along the underlying hierarchy when contributing to the ontology. Given the observations made for ICD-11 this behavior can be enforced by providing certain functionalities in the user-interface especially when they compliment the workflows of the contributors.

The results of this analysis have also shown that users are likely to pursue a certain strategy or intermediate goal for their edit sessions, for example changing all classes in a specific (narrow) area of the ontology. This is evident in the observation that after returning from a \emph{BREAK}, users have a very high tendency to change the ontology ``somewhere else'' (see the transition probabilities from \emph{BREAK} towards \emph{Other} in the top-row of Figure~\ref{fig:rsuc}), rather than picking up the work, where they left off.
This discovery is very important for developing class-recommender, as we may use the results of this analysis to suggest closely related classes to the current class a user is working on, however when that user stays inactive for the duration defined for introducing \emph{BREAK}s the recommendation strategy has to be changed.

\subsection{Property paths}
\label{sub:content and user-interface paths}

Aside from analyzing different aspects of activity (Section~\ref{sub:activity paths}) and the correlation between contribution patterns and the structure of an ontology (Section~\ref{sub:structural paths}), we can use Markov chains to perform an analysis on the properties that are consecutively change by users in an ontology. \green{This means that, for example, if a property value was edited by a user, we extracted the property (not the value) and created chronologically ordered lists of properties, whose values were changed by the corresponding users. For example, if a user changed the title of a specific class, we would extract \emph{title}, rather than the value inserted into the title property.} Now, we provide insights into emerging patterns from different viewing angles for the observations. Thus, we look at property sequences for (a) single users (user-based) and for (b) single classes (class-based) -- see Section~\ref{sub:sequential paths}. We were not able to perform the \emph{Property Paths} analysis on OPL and BRO as these datasets contain only a very limited number of unique property \green{value} changes during our observation periods. We also had to discard the results from NCIt, as the ontology-editing environment for NCIt provides a unique change-queuing mechanism that allows for multiple property \green{values} to be changed at the same time, making it impossible to extract chronologically ordered sequential property patterns.

\begin{figure*}[!ht]
\centering
\subfigure[\captionicd (Class)]{\label{fig:plcs:a}\includegraphics[width=\subfiguresizecontent\textwidth]{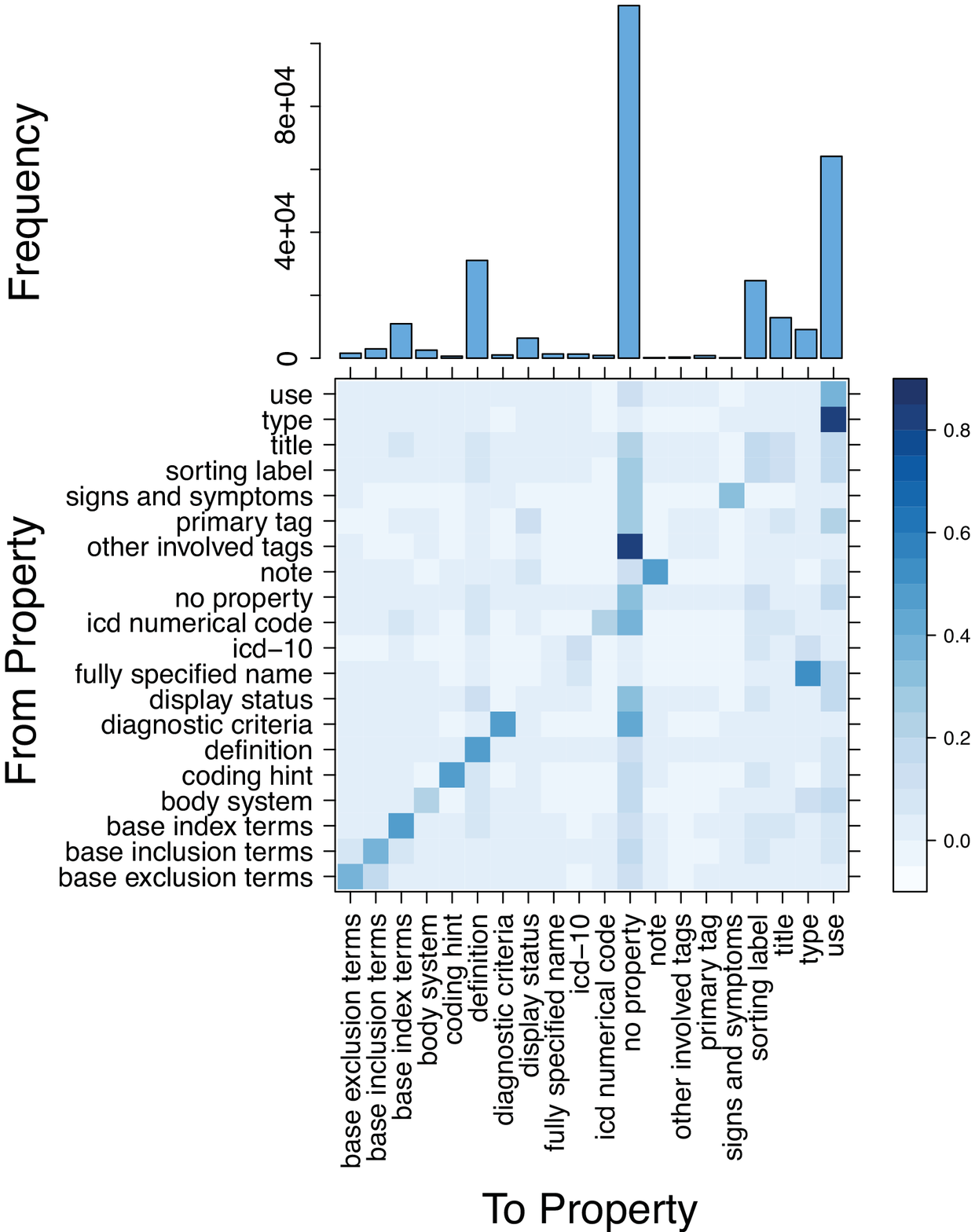}}
\subfigure[\captionictm (Class)]{\label{fig:plcs:c}\includegraphics[width=\subfiguresizecontent\textwidth]{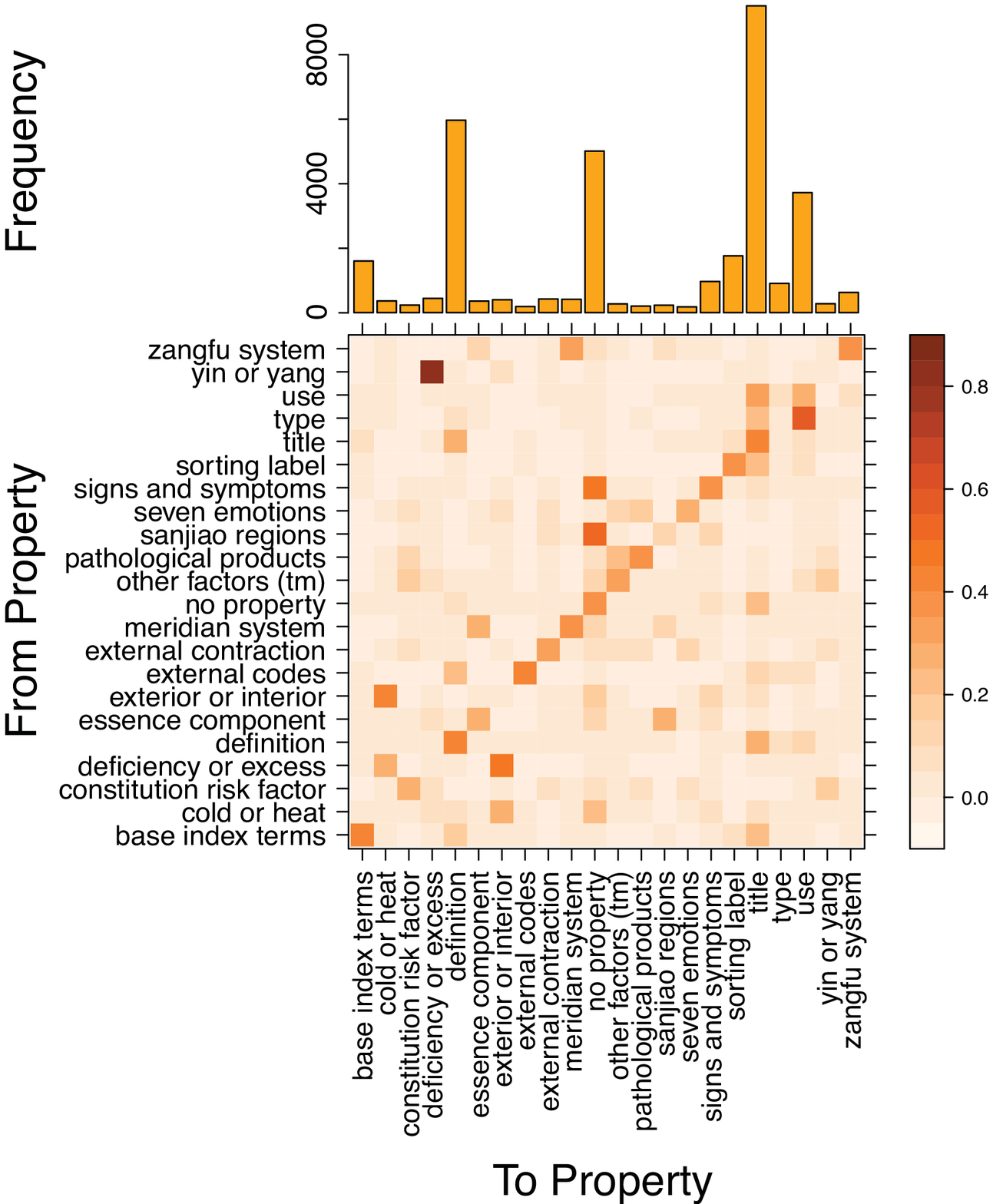}}\\
\subfigure[\captionicd (User)]{\label{fig:plcs:b}\includegraphics[width=\subfiguresizecontent\textwidth]{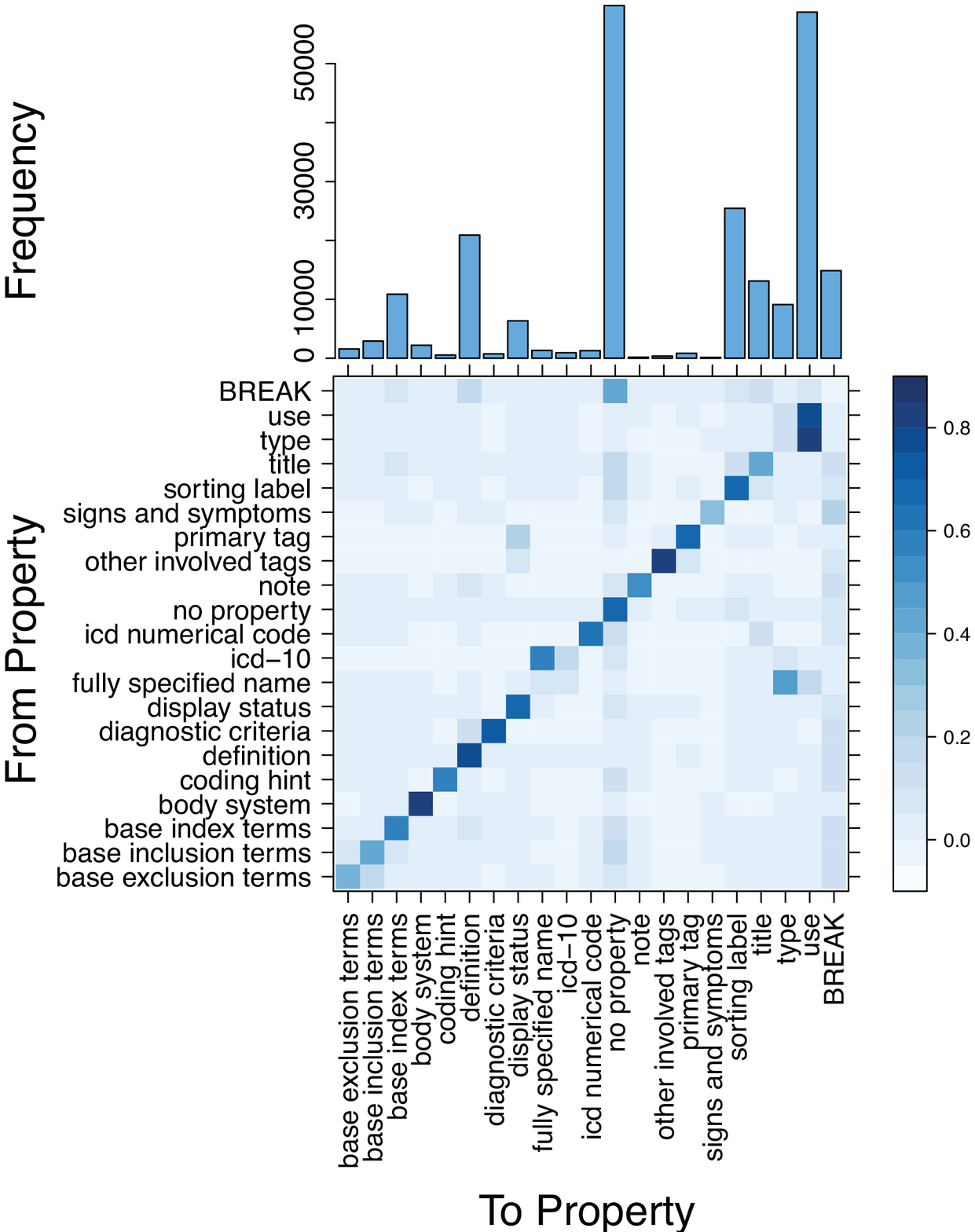}}
\subfigure[\captionictm (User)]{\label{fig:plcs:d}\includegraphics[width=\subfiguresizecontent\textwidth]{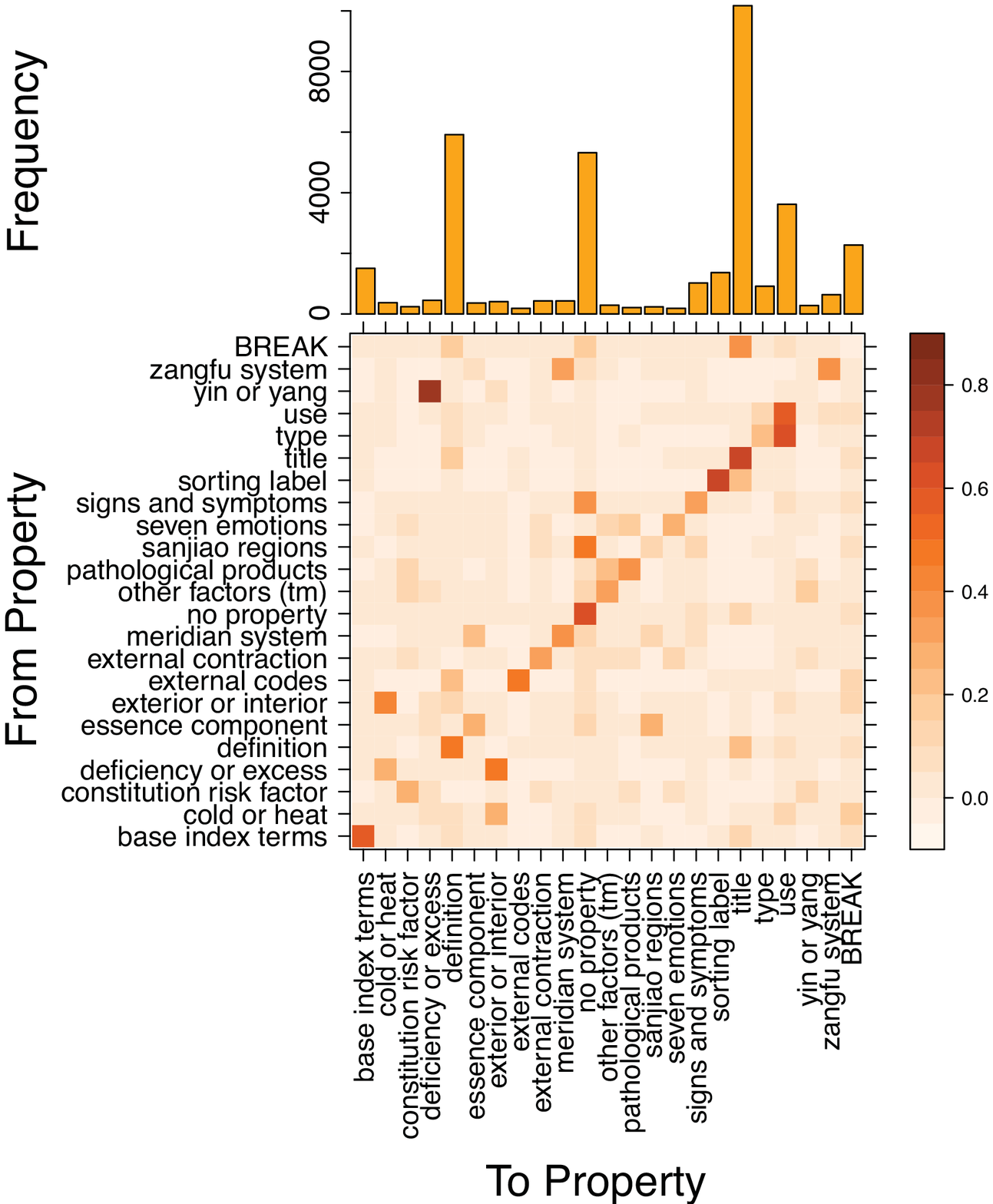}}
\caption{\textbf{Results for the \emph{Property Paths} analysis:} The columns and rows of the transition maps (\textbf{bottom area} of Figures~\ref{fig:plcs:a} to \ref{fig:plcs:d}) represent the transition-probabilities of a first-order Markov chain between consecutively changed properties, where rows are \emph{source properties} and columns are \emph{target properties}. Figures~\ref{fig:plcs:a} and \ref{fig:plcs:b} represent class-based patterns while Figures~\ref{fig:plcs:c} and \ref{fig:plcs:d} visualize user-based patterns.
A sequence (or transition-probability) is always read \emph{from row to column}. Darker colors represent higher transition-probabilities while lighter colors indicate lesser transition-probabilities. Absolute probability values are dependent on the number of investigated rows and columns, hence relative differences are of greater importance. Across all datasets a very clear trend towards consecutively editing the same properties can be observed.
The histograms (\textbf{top area} of Figures~\ref{fig:plcs:a} to \ref{fig:plcs:d}) show the total edits of each property in the corresponding datasets aggregated over all users and classes (again for a first-order Markov chain).
Note, that the $y$-axes for all histograms are scaled differently for each dataset. As ICTM and ICD-11 only share a limited amount of properties the $x$-axes (and column/rows of the transition maps) are different from project to project. In both projects and across all 4 different approaches the \emph{title}, \emph{definition} and \emph{use} properties are frequently used. Due to reasons of readability we were forced to remove properties from the plots, which exhibited only a very limited number of changes, thus did not provide substantial information for the purpose of this analysis.}
\label{fig:plcs}
\end{figure*}

\green{
\textbf{Path \& model description:} First, we extracted the properties whose values were changed in ICD-11 and ICTM, sorted either by user and timestamp or by class and timestamp. Finally, two different types of chronologically ordered property lists were extracted, one ordered per user and one ordered per class (for both datasets). 
The properties in \emph{Property Paths} represent the ones which can be assigned a value for each class in ICD-11 and ICTM.
Whenever a change did not modify a property (e.g., because the change action dealt with moving or creating a class) we added the element \emph{no property} to the corresponding path.
A potential path for a single user or class then may look like: \emph{title, title, title, use}.
 Similar to previous analyses, if the same user has consecutively changed the same property (e.g., in the previous example \emph{title}) on the same class, we  merged these multiple changes into one successive change. Analogously, however without the restriction of the same user, if the same property was changed on the same class, we merged these changes into one sequent change. For previous example, if changes would have been performed editing the referenced properties for a single class, we would end up with the path: \emph{title, title, use}.

Consequently, we fit a first-order Markov chain model on this set of paths (for users or classes). The final transition probabilities of the model then give us information about the probability of changing a value of one property Y after another property X either for users or for classes. For instance, we can find the property Y that most frequently has been changed after property X for classes.
}

\textbf{Results:} When looking at the histograms (top area in Figures~\ref{fig:plcs:a} to \ref{fig:plcs:d}) we can see that even after removing not very frequently used properties,\footnote{All properties which where rarely edited have been removed from Figure~\ref{fig:plcs} as they do not hold information but their removal increased the readability of the plots dramatically.} both datasets exhibit a few properties which have received a high number of changes, while the remaining majority of properties only received a very limited number of changes. For both datasets, aside from \emph{no property}, the properties \emph{use}, \emph{title} and \emph{definition} appear to be the most frequently used properties.
As can be seen in the top area of Figures~\ref{fig:plcs:a} and \ref{fig:plcs:c},  multiple consecutive changes of the same property appear to be fairly common for both datasets. In contrast, when looking at Figures~\ref{fig:plcs:b} and \ref{fig:plcs:d}, which depict the transition probabilities between the sequences of properties changed by each user, we can see an even stronger trend towards consecutively changing the same properties across different classes, especially \emph{definition}, \emph{title} and \emph{use}. For ICD-11 Figures~\ref{fig:plcs:a} and \ref{fig:plcs:b} show that the class-based approach is less focused on consecutively changing the same property, evident in the brighter diagonal, when compared to the user-based approach. This is due to the export functionality available in iCAT combined with the manual process of inserting the same property for different classes by users of ICD-11. In contrast, such functionality is absent in ICTM, thus leading to similar behaviors for the class and user-based approaches for ICTM. The fact that a large portion of successive changes are conducted on the same property for both approaches analyzed for ICTM could also be due to the multilingual nature of the project, meaning that certain properties, such as \emph{title} and \emph{definition}, have to be entered multiple times in multiple languages. Similar results have been presented by \citet{wang2013analysis}, who used association rule mining techniques to analyze the change-logs of ICD-11 and ICTM.

Contributors in ICD-11 have a high tendency of performing \emph{no property} changes after they return from a \emph{BREAK} followed by \emph{use}, \emph{title} and \emph{definition}. In ICTM,  users resume their work primarily by changing the \emph{title} property, the \emph{definition} property followed by \emph{no property} changes.

\textbf{Interpretation \& practical implications:} One of the main benefits of this analysis is the identification of commonly and consecutively changed properties for classes and users. In turn, this information might potentially be used to suggest work (e.g., prompting a user to check a certain property by combining the \emph{User-Sequence Paths} analysis and the \emph{Property Paths} analysis), or by ontology-engineering tool developers to potentially anticipate the property a user is most likely to change next. The fact that classes appear to exhibit more diverse property-contribution patterns when being changed than users could be a direct result of the multi-lingual nature of ICTM and the already mentioned export functionality present in iCAT.
This means that given the most recent property of a class that was edited, we may predict which property is most likely to be changed next. Similarly, we can predict the property a user is going to edit next.

\begin{table*}[t!]
\center
\footnotesize
\caption{\textbf{A summary of all findings} applicable to all investigated biomedical ontologies. All listed findings are discussed in more detail in Section~\ref{discussion}. }
\begin{tabular}{| c | p{6.5cm} | p{7.5cm} |}\hline
\multirow{5}{*}{\shortstack{User-sequence paths\\(cf. Section~\ref{sub:activity paths})}} & \multirow{2}{*}{\textbf{Users work in micro-workflows}} & Information about 
which users successively change a class can be identified; i.e., information about
who has edited classes in the past contains predictive information about who is going to change a class next.\\\cline{2-3}
& \multirow{4}{*}{\textbf{User-roles can be identified}} & Looking at historic data, we can identify different user roles, i.e., administrators and moderators, gardeners \green{(a contributor focused on pruning ontology classes and fixing syntactical errors)} and users that frequently interact with (collaborate/revert) each other. \\
\hline\hline
\multirow{9}{*}{\shortstack{Structural paths\\(cf. Section~\ref{sub:structural paths})}} & \multirow{2}{*}{\textbf{Users' edit behavior is influenced by the class hierarchy}} & Contributors, when adding content to the ontology, are influenced by the class hierarchy.\\\cline{2-3}
& \multirow{5}{*}{\textbf{Users edit the ontology top-down and breadth-first}} & By and large, users exhibit a minor tendency towards top-down editing behavior when changing hierarchy levels while contributing. However, when staying in the same hierarchy level, contributors rather follow a \emph{breadth-first} edit behavior, moving from one sibling of a class to the next sibling.\\\cline{2-3}
& \multirow{3}{*}{\textbf{Users edit closely related classes}} & Contributors have a very high tendency to consecutively change closely related classes, as opposed to randomly and distantly related classes. \\\hline\hline
\multirow{3}{*}{\shortstack{Property paths\\(cf. Section~\ref{sub:content and user-interface paths})}} & \multirow{3}{*}{\textbf{Users perform property-based workflows}} & Contributors, when adding content to the ontology, tend to concentrate their efforts on one single property, which is added and edited for multiple classes.\\\hline
\end{tabular}
\label{tab:findings summary}
\end{table*}

\section{Findings and discussion}
\label{discussion}

In this section we first summarize our findings in Section~\ref{sub:summary}
before we
shortly discuss the potential applicability of higher order Markov chain models in Section~\ref{sub:order}. Next, we discuss differences between the investigated projects in Section~\ref{sub:differences} and finally, point out potential limitations of this work in Section~\ref{sub:limitations}.

\subsection{Summary of findings}
\label{sub:summary}

We will now discuss our main findings (Table~\ref{tab:findings summary}) and explore their consequences. 

\textbf{Emergence of micro-workflows:} By investigating whether sequential user-contribution patterns (see Section~\ref{sub:activity paths}) can be identified in \numdatasets different collaborative ontology-engineering projects, we have shown that users appear to work in micro-workflows, indicating that for all investigated projects, each user contains predictive information about the user, who is going to contribute to a specific class next.

Additionally, however not presented in this paper due to reasons of space, we have also conducted an analysis to determine the change type (e.g., adding a property value, moving a class, replacing a property value, etc.) a user is most likely to perform next (as shown in \citet{walk_dev_mc} for ICD-11). In this analysis we were able to extract a first-order Markov chain for all datasets presented in this paper, meaning that the last change type that a user performed contains information about the next change type of that user. When combining the information about the user who is most likely to contribute to a class next and the specific change action that this user is most likely to conduct (or  the change action that is most likely conducted on a class next), we can create specific tasks for contributors, asking them to perform a certain change on a specific class.

Our results \green{could} be used by project managers and ontology-engineering tool developers to identify classes for users and users for classes, helping editors to minimize the necessary efforts for finding and identifying classes to contribute to.
Moreover, automatic means of curating and delegating work-tasks to users can be derived by ontology-engineering tool developers, which can help to potentially increase participation as discussed in \citet{kittur2008}.

\textbf{User roles can be identified:} Across all datasets we were able to identify that a limited number of users have contributed to the majority of all changes. These highly active users are very likely to be \emph{target users} for all other users, meaning that they are very likely to change the same class after another user. Across all \numdatasets datasets, the roles of these \emph{target users} could be identified by us as moderators or administrators of the corresponding projects performing maintenance tasks, such as gardening (e.g., pruning outdated classes, fixing errors, etc.) or manual verification of newly added data.

Furthermore, we were able to show that moderators and administrators divide work among each other, as they are not very likely to change the same classes directly after another administrator or moderator, even though these users exhibit the highest absolute numbers of changes in the corresponding projects. Looking at the transition probabilities of Figure~\ref{fig:ucsl} it is possible to identify users or even groups of users who have a high tendency to work on the same classes, thus might be collaborators or reverting/correcting changes of each other.

\textbf{Users edit the ontology top-down and breadth-first:} The \emph{Depth-Level Paths} analysis (see Section~\ref{subsub:depth-level paths}) demonstrated that users have a very high tendency of staying in the same depth level when contributing to the ontology. If editors change depth levels while editing the ontology they exhibit a minimal preference to do so in a \emph{top-down} rather than a \emph{bottom-up} manner. Furthermore, the results suggest that users move along the hierarchy as we were able to show that they follow a \emph{top-down} editing strategy for classes that are closer to the root node while this changes to a \emph{bottom-up} editing strategy for classes closer to the deepest depth levels and transitions are more likely to occur along the immediate higher or lower depth level. 

To further investigate the distances between changed classes at the same depth levels we investigated the \emph{Hierarchical Relationship Paths} (e.g., child, parent, sibling, cousin, etc.) between these changed classes. We found  that users, when they edit classes on the same depth level, follow a \emph{breadth-first} manner, focusing on editing all the siblings of a class before switching to a completely different area of the ontology to continue their work after a \emph{BREAK}. 

\textbf{Users edit closely related classes:} \green{Additionally to the \emph{breadth-first} manner that users follow when editing classes in the same depth level, we discovered that users have a very high tendency to work on closely related classes (e.g., the sibling or cousin of the currently changed class).}
The information collected in Section~\ref{sub:structural paths} \green{allows to potentially} predict (or narrow down) the class a user is going to contribute to next, which\green{, if accurate}, is a very valuable information that could be used for a variety of improvements and adaptions. \green{For example, project-administrators could adjust the milestones of the development-strategy to better reflect the way users contribute to the ontology while user-interface designers could emphasize certain areas of the ontology to direct users towards specific classes -- especially after they return from a \emph{BREAK} -- or implement pre-fetching algorithms to minimize load-times.}
For contributors in particular, the task of identifying and finding classes that they (i) want and (ii) have the necessary expert knowledge to contribute to is a time-consuming task, which potentially can be minimized by implementing class recommender based on the results of the \emph{Structural Paths Analysis} and \emph{User-Sequence Paths Analysis}.

\textbf{Users perform property-based workflows:} The investigation of sequential patterns for property-contributions showed that in ICD-11, users have a very high tendency of consecutively changing the same property across multiple classes. 
We could also identify specific patterns that emerge when users successively change properties in collaborative ontology-engineering projects.

The results collected in the Section~\ref{sub:content and user-interface paths} provide new insights for administrators and ontology-engineering tool developers, as they allow the generation of work-tasks (e.g., Please verify the property \emph{title} of the class \emph{XII Diseases of the skin}!). So far, users are always presented first with the section of the interface that allows for changing or adding the \emph{title} and \emph{definition}, which could be one explanation for the high probabilities of users changing these properties when returning from a \emph{BREAK}.

Note, that for this analysis we have used the data from ICD-11 and ICTM, which both share a very similar ontology-engineering tool, thus the results might be biased towards the used ontology-editor.

\green{
\subsection{Higher order Markov chains}
\label{sub:order}

Based on our proposed methodology of using first-order Markov chain models (see Section~\ref{sub:markov chain model selection}) resulting in the findings summarized in Section~\ref{sub:summary}, we currently lay our focus on detecting patterns only derived from successive interactions within collaborative ontology-engineering projects. This means, that we identify how likely it is that one specific interaction follows another one (e.g., which user edits a class after another one). This is reasoned by the definition of a first-order Markov chain based on the Markovian property which postulates that the next interaction only depends on the current one.

Contrary, Markov chain models can also be defined on higher orders; this means that the next state of the model (or interaction in our case) depends on a series of preceding ones instead of only the current one. For example, a 
\emph{second-order} Markov chain model postulates that the next state depends on the current state and also the previous one. Previous studies suggest that human navigation on the Web might be better modeled by using higher order models compared to first-order models (e.g., \cite{chierichetti,singer_mc}). Hence, we could assume that this might also be the case for our use-case.
By also modeling our data with such higher order models, we would potentially be able to identify longer patterns (e.g., \emph{User A} regularly edits a class after \emph{User B and User C}). Also, possible recommender systems could benefit from the additional predictive power of such higher order chains.\footnote{Note that it is necessary to apply model selection techniques as described in \cite{singer_mc} in order to identify the most appropriate Markov chain order based on statistical significant improvements of higher orders compared to lower orders} While highly interesting, this analyses would be out-of-scope for this article which is why we leave this open for future work.
}

\subsection{Differences between the investigated projects} 
\label{sub:differences}

Even though each project exhibits a different number of depth levels, which all receive a different amount of attention by the contributors, we can observe commonalities of edit strategies between them. For example, the levels $3$ to $6$ exhibit the highest number of changes in our observation period for ICD-11, while for OPL these levels are $6$ and $7$.

Regarding the hierarchical relationships we can see that consecutively changing the same class is very likely to happen in ICD-11, ICTM, BRO and OPL regardless of the source relationship (evident in the darker colored \emph{Self} columns in Figures~\ref{fig:rsuc:a}, \ref{fig:rsuc:b}, \ref{fig:rsuc:d} and \ref{fig:rsuc:e}). This \emph{Self}-relationship is still very prominent, however the transition probabilities towards \emph{Self} for NCIt are not as dominant as they are for the other datasets.

Another observation depicted in the transition maps is the clear focus on transitions from \emph{Sibling} to \emph{Sibling} across three out of five datasets, with the exception of ICTM and OPL. One explanation for ICTM could be the fact that some properties of the ontology are multi-lingual, thus require users to add multiple languages for the same property, which are all stored as a single change. For OPL, transitions, except towards \emph{Self} are in general really scarce, indicating that users focused on editing and entering multiple property values (or one property value) of a single class before continuing to the next class.

When looking at the sequence of changed properties for each class (in contrast to: for each user) we can observe a concentration on consecutively changing the same property in ICTM, which is most likely a direct result of the multi-lingual nature of the properties used in this project. In ICD-11 on the other hand, transitions between changed properties of classes are much more diverse and less focused on transitions between the same properties. This observation indicates that either not all properties have received a substantial amount of values for all the possible properties and/or that users make use of this special export functionality of iCAT, thus successively changing the same property is less common as the content is only inserted once into the system.

In the \emph{User-Interface Sections Paths} analysis we have mapped the changed properties to the corresponding sections of the user interface of the used ontology-engineering tools, which essentially represents a more abstract analysis of the \emph{Property Paths} analysis. By investigating the sequences of user interface sections we could confirm that, for ICD-11, users have a very high tendency to consecutively change the same properties for multiple classes, evident in the scarce transitions between different sections and the high concentration on transitions between the same sections. For ICTM this behavior was not as distinctive as it was for ICD-11, which could be due to the missing export functionality and therefore the lack of the previously explained manual import sessions.

In general these observations indicate that the absence or presence of a given functionality of the ontology-engineering tool can produce (and influence) different editing behaviors when developing an ontology.

\subsection{Limitations}
\label{sub:limitations}

We were not able to recreate the exact class hierarchy of the ontology for every single change across our observation periods for all datasets. This limitation is partly due to a lack of detail in the change-logs. Thus, we decided to focus our analysis, using all \numdatasets ontologies \emph{as is} at the latest point in time, which is also what would most likely be used in a \emph{real-world} scenario.

For example, if a class was changed by a user while it was located on depth level $3$ and at a later point in time moved to a different location where it now resides at depth level $5$, we would assume that this class has always been on depth level $5$. Please note that this bias is only present in the \emph{Structural Paths} analyses (Section~\ref{sub:structural paths}). To measure the extent of the potential bias, we counted all changes that were performed on a class before it was moved within in the ontology. Applying this rule to our change dataset, we collected a total of $116,204$ of $439,229$ changes for ICD-11 and $18,958$ of $67,522$ for ICTM. These numbers represent about $1/4$ and $1/3$ of all changes for ICD-11 and ICTM respectively. For BRO $276$ of $2,507$ (ca. $1/10$) and for OPL $2$ of $1,993$ of all changes were performed on classes, which have been moved afterwards.

Note  that an additional requirement for the identification of sequential patterns in collaborative ontology-engineering projects using Markov chains is the availability of rather large change-logs. In general, the less common entities (e.g., properties) are present in the change-log the more (exponentially) observations have to be available in order to detect more fine-grained patterns. Without enough observations (changes), the identification of sequential patterns is either very hard, and can only be approximated, or not possible at all. As can be seen in Table~\ref{tab:dataset details}, we have selected all of our datasets to satisfy this requirement, as all chosen datasets exhibit a substantial number of changes.

Furthermore, we have included \emph{artificial session breaks} into our analysis as described by \citet{walk_dev_mc} to analyze where or what users start to edit in the ontology and where or what users edit before they take a break. For all user-based analyses we have introduced a \emph{BREAK} if two consecutive changes of the same user were apart longer than $5$ minutes.

\green{All analyses in this paper are based on \emph{isKindOf} relationships for determining distances and locations within the ontology. We plan on further expanding this analysis by investigating the impact of other kinds of relationships and other features that are available in ontologies on our pattern detection approach.}

\green{Even though all datasets presented in this paper are created with \wpro or one of its derivatives, there is only one requirement that prevents practitioners from performing this analysis on other ontologies: The availability of a change-log (in the required granularity for the deemed analyses) that can be mapped onto the underlying ontology. Note that it would be possible to conduct this analysis for ontologies created by single individuals, meaning that ``collaboration'' is only a requirement when the nature of the analysis requires investigating transitions between multiple users. 

Also, the kind of knowledge base (classification, taxonomy or ontology), the used representation language (e.g., OWL and OWL-DL expressivity, RDF, Turtle) or the development tool of a particular collaborative ontology-engineering project in question does not prohibit conducting a pattern analysis as presented in this paper, as long as the underlying knowledge base (and thus the change-log) exhibits the necessary granularity and the semantic properties of interest for the analysis.}

\green{However, this also means that the differences of the knowledge representation used languages (i.e., expressivity and types) are not considered by our analysis, with NCIt being a thesaurus and the rest of the investigated datasets being ontologies. Thus, whenever differences are observed between NCIt and the remaining datasets, further research is warranted to determine the origin of this observation.}

\green{Furthermore, the analysis presented relies on investigating usage logs of collaborative ontology-engineering projects by looking at changes, performed by users of the corresponding systems. As this only represents one possible way of interacting with the underlying ontology, albeit the most frequently used one, an extension of the conducted Markov chain investigation warrants future work to include, for example, discussions for consensus building, suggestions of terms by users or automatic imports.}

\section{Related work}
\label{related work}

For the analysis and evaluation conducted in this paper, we identified relevant information and publications in the domains of (i) Markov chain models, (ii) collaborative authoring systems and (iii) sequential pattern mining.

\subsection{Markov chain models}
\label{sub:markov chain models}

In the past, Markov chain models have been heavily applied for modeling Web navigation -- some sample applications of Markov chains can be found in \cite{borges, deshpande, lempel, pirolli,sen, zukerman}. 
Also, the Random Surfer model in Google's PageRank \cite{brin} can be seen as a special case of a Markov chain.

Previously, researchers  investigated whether human navigation is memoryless (i.e., of first order) in a series of studies (e.g., \cite{borges1999, pirolli}). However, these studies mostly showed that the memoryless model seems to be a quite plausible abstraction (see e.g., \cite{cadez, sarukkai, sen, zukerman}).
Recently, a study picked up on these investigations and suggested that the Markovian assumption (i.e., property) might be wrong \cite{chierichetti}. However, this study did not reveal any statistically significant improvements of higher order models. \citet{singer_mc} solved this problem by developing a framework for determining the appropriate order of a Markov chain for a given set of input data. 
In \citet{walk_dev_mc} we applied and mapped the presented framework onto structured logs of changes and provided an in-depth description of the requirements and steps necessary to use the framework in this setting.

In this paper we present a detailed analysis of sequential patterns by applying and analyzing Markov chains across the change-logs of \numdatasets collaborative ontology-engineering projects in the biomedical domain. A more detailed explanation of the necessary steps to be able to apply Markov chains onto the change-logs of collaborative ontology-engineering projects is presented in \citet{walk_dev_mc}. Note that we focus on applying first-order Markov chain models in this work while we see the application of also higher order models as highly interesting future work as discussed in Section~\ref{sub:order}.

\subsection{Collaborative authoring systems}
\label{sub:collaborative authoring systems}

Research on collaborative authoring systems such as Wikipedia has in part focused on developing methods and studying factors that improve article quality or increase user participation. These problems represent important facets of collaborative authoring systems and solutions to tackle these problems are of interest for collaborative ontology-engineering projects.

For example, \citet{cabrera2002} demonstrated the effect of minimizing the costs and efforts necessary for users to contribute on potentially achieving  higher contribution rates. 
Another approach, also presented by \citet{cabrera2002}, focuses on providing an environment where interactions and communication between contributors are encouraged and performed frequently over a long period of time to establish a group identity and to promote personal responsibility. 

More recent research on collaborative authoring systems, such as Wikipedia, focuses on describing and defining not only   the act of collaboration amongst strangers and uncertain situations that contribute to a digital good \cite{conf/wikis/KeeganGC11} but also on antagonism and sabotage of said systems \cite{shachaf2010beyond}.
It has also been discovered only recently that Wikipedia editors are slowly but steadily declining \cite{Suh09singularity}. Therefore \citet{conf/wikis/HalfakerKR11} have analyzed what impact reverts have on new editors of Wikipedia. \citet{kittur2008} showed that an increase in participation can be achieved by directly delegating specific  tasks to contributors. As simple as this approach may appear, the identification of work (and thus specific  tasks) is still a tedious and time-consuming process, which can only partly be automated due to its assigned complexity.

With the analysis that we described here, we provide new  results that we can   use to tackle some of the  problems for collaborative authoring systems. These problems are also present in collaborative ontology-engineering projects. For example, we can identify new tasks
 by combining the results of the \emph{User-Sequence Paths} (Section~\ref{sub:activity paths}) and \emph{Property Paths} (Section~\ref{sub:content and user-interface paths}) analyses to suggest classes and the corresponding properties to work on to users.

\subsection{Sequential pattern mining}
\label{sub:sequential pattern mining}

In 1995 \citet{Agrawal:1995:MSP:645480.655281} have first addressed the problem of sequential pattern mining. They stated that given a collection of chronologically ordered sequences, sequential pattern mining is about discovering all sequential patterns weighted according to the number of sequences that contain these patterns. The presented algorithm represents one of the first \emph{a priori} sequential pattern mining algorithms. This means that a specific pattern cannot occur more frequently (above a threshold) if a sub-pattern of this pattern occurs less often (below that threshold). Other examples of a priori algorithms are \cite{Ng:1998, Sarawagi:1998}.

One of the biggest problems assigned to the a priori based sequential pattern mining algorithms was (in the worst case) the exponential number of candidate generation. To tackle this problem \citet{Han:2000:MFP:342009.335372} developed the FP-Growth algorithm.

Many researchers have adapted different algorithms and approaches for different domains to anticipate changing requirements, such as \citet{Wang:2004:BEM:977401.978142} and \citet{hsu2007identification} who analyzed algorithms for sequential pattern mining in the biomedical domain.

In \citet{walk_dev_mc} the authors have presented a novel application of Markov chains to mine and determine sequential patterns from the structured logs of changes of collaborative ontology-engineering projects.
Making use of this framework we investigate differences and commonalities across \numdatasets different collaborative ontology-engineering projects from the biomedical domain.

\section{Conclusions \& future work}
\label{conclusions and future work}

In this work, we discovered intriguing social and sequential patterns that suggest that large collaborative ontology-engineering projects are governed by a few general principles that determine and drive development. Specifically, our results indicate that patterns can be found in all investigated projects, even though the \captionncit, the \captionicd, the \captionictm, the \captionopl and the \captionbro (i) represent  different projects with different goals, (ii) use variations of the same ontology-editors and tools for the engineering process and (iii) differ in the way the projects are coordinated. \green{Using the presented Markov chain analysis, multiple different user-roles could be identified in all investigated datasets. We were also able to see that users work in micro-workflows, meaning that given a specific user, we can identify the most likely users that are editing a specific class next, again independent from the investigated project. When contributing to a project that is created using \wpro, \icat, iCAT-TM or \cprot, users exhibit a tendency to do so in a \emph{top-down} and \emph{breadth-first} manner, editing primarily closely related classes while moving along the ontological hierarchy. In ICD-11 and ICTM we were able to identify property-based workflows, meaning that users concentrate their efforts on adding and editing values for one specific property for multiple classes.}

The analysis presented not only provides new insights about the engineering and development processes of each single project, but also shows that the analysis of sequential patterns \green{potentially provides} actionable insights for different stakeholders in collaborative ontology-engineering projects.

Furthermore, the information of the next possible action (e.g., a user, a change-type, a property, set of classes) or the combination of multiple of these next actions \green{could} be used by ontology-engineering tool developers to \green{potentially} augment users in collaboratively creating an ontology. For example, by \green{making use of the \emph{Property Paths} analysis to} highlight, prefetch, rearrange or adjust sections and content of the interface dynamically, according to the user's needs.

\green{The next logical step to further deepen our understanding of collaborative ontology-engineering projects involves applying the gathered results to productive and live environments, for example as plug-in for (Web)Prot\'eg\'e. Simultaneously, this would allow us to collect valuable data to quantify the usefulness and actionability of the results, generated with our presented approach, in real world scenarios.} 

\green{Additionally, expanding the Markov chain analysis to take other types of interactions (e.g., discussions, automatic imports and term suggestions by users) into account, represents a potential topic of future work. This also includes a detailed analysis of human factors studies in terms of user-studies (e.g., with a heuristic evaluation or A/B testing) or more sophisticated approaches, such as eye tracking, to assess the usefulness of the presented results for augmenting users when collaboratively engineering an ontology.}

Furthermore, as change tracking and click tracking data will \green{likely} become available more broadly \green{in the future}, we believe that the analysis of this paper and the possible benefits of putting the results into practical use represent an import step towards the development of better (and simpler) ontology editors, which can dynamically anticipate the editing-style of the users. Project administrators could make use of the results of the analysis, for example by allowing for easier delegation of work to the ``right'' users. This is even more emphasized when considering that the Markov chain analysis is not computationally intensive, making it highly suitable for productive use.

As biomedical ontologies play an increasingly critical role in acquiring, representing, and processing information about human health, we can use quantitative analysis of editing behavior to generate \green{potentially} useful insights for building better tools and infrastructures to support these tasks.

\section*{Acknowledgement}

\small
This work was generously funded by a Marshall Plan Scholarship with support from Graz University of Technology. Further, this work is supported in part by grants GM086587 and GM103316 from U.S. National Institutes of Health.

\bibliographystyle{elsarticle-num-names}

\end{document}